\documentclass[11pt]{article}
\usepackage{axodraw}

\renewcommand{\theequation}{\arabic{section}.\arabic{equation}}
\def\beq{\begin{eqnarray}}
\def\eeq{\end{eqnarray}}
\def\bea{\begin{eqnarray}}
\def\eea{\end{eqnarray}}
    
\newcommand{\newc}{\newcommand}
\newc{\gsim}{\lower.7ex\hbox{$\;\stackrel{\textstyle>}{\sim}\;$}}
\newc{\lsim}{\lower.7ex\hbox{$\;\stackrel{\textstyle<}{\sim}\;$}}
\newc{\lnbar}{\overline{\rm ln}}
\newc{\dilog}{{\rm Li}_2}
\newc{\trilog}{{\rm Li}_3}
\newc{\cTarasov}{a}
\newc{\cCCLR}{b}
\newc{\cPassarino}{c}

\catcode`@=11
\long\def\@caption#1[#2]#3{\par\addcontentsline{\csname
  ext@#1\endcsname}{#1}{\protect\numberline{\csname
  the#1\endcsname}{\ignorespaces #2}}\begingroup
    \small
    \@parboxrestore
    \@makecaption{\csname fnum@#1\endcsname}{\ignorespaces #3}\par
  \endgroup}
\catcode`@=12

\begin{document}
\begin{titlepage}

\begin{flushright}
hep-ph/0501132
\end{flushright}

\vspace{0.3in}

\begin{center}
{\large\bf {\tt TSIL}: a program for the calculation of two-loop
self-energy integrals}
\end{center}

\vspace{.15in}

\begin{center}
{\sc Stephen P.~Martin$^{a,b}$ and David G. Robertson$^c$}

\vspace{.1in}

{(a) \it Department of Physics, Northern Illinois University, 
DeKalb IL 60115}\\
{(b) \it Fermi National Accelerator Laboratory,
P.O. Box 500, Batavia IL 60510}\\ 
{(c) \it Department of Physics and Astronomy,
Otterbein College, Westerville OH 43081 }\\ 
\end{center}

\vspace{0.5in}

\begin{abstract}
\noindent {\tt TSIL} is a library of utilities for the numerical
calculation of dimensionally regularized two-loop self-energy integrals. A
convenient basis for these functions is given by the integrals obtained at
the end of O.V.~Tarasov's recurrence relation algorithm.  The program
computes the values of all of these basis functions, for arbitrary input
masses and external momentum.  When analytical expressions in terms of
polylogarithms are available, they are used. Otherwise, the evaluation
proceeds by a Runge-Kutta integration of the coupled first-order
differential equations for the basis integrals, using the external
momentum invariant as the independent variable. The starting point of the
integration is provided by known analytic expressions at (or near) zero
external momentum. The code is written in C, and may be linked from C, C++,
or Fortran.  A Fortran interface is provided.  We describe the structure
and usage of the program, and provide a simple example application. 
We also compute two new cases analytically, 
and compare all of our notations and conventions for the two-loop 
self-energy integrals to those used by several other groups. 

\end{abstract}

\end{titlepage}
\baselineskip=17pt
\setcounter{footnote}{1}
\setcounter{page}{2}
\setcounter{figure}{0}
\setcounter{table}{0}

\tableofcontents

\section{Program summary}
\label{sec:summary}
\setcounter{equation}{0}
\setcounter{footnote}{1}

\noindent
{\it Title of program:} {\tt TSIL}

\vspace{0.35cm}

\noindent
{\it Version number:} 1.4  

\vspace{0.35cm}

\noindent
{\it Available at: }{\tt 
http://www.niu.edu/spmartin/TSIL/}\\
\phantom{{\it Available at: }}{\tt 
http://faculty.otterbein.edu/DRobertson/tsil/}

\vspace{0.35cm}

\noindent
{\it Programming Language:} C

\vspace{0.35cm}

\noindent
{\it Platform:} Any platform supporting the GNU Compiler Collection (gcc),
the Intel C compiler (icc), or a similar C compiler with support for complex
mathematics

\vspace{0.35cm}

\noindent
{\it Distribution format:} ASCII source code

\vspace{0.35cm}

\noindent
{\it Keywords:} quantum field theory, Feynman integrals, two-loop integrals, 
self-energy corrections, dimensional regularization

\vspace{0.35cm}

\noindent
{\it Nature of physical problem:} Numerical evaluation of
dimensionally regulated Feynman integrals needed in two-loop self-energy 
calculations in relativistic quantum field theory in four dimensions

\vspace{0.35cm}

\noindent
{\it Method of solution:} Analytical evaluation in terms of
polylogarithms when possible, otherwise through Runge-Kutta 
solution of differential equations

\vspace{0.35cm}

\noindent
{\it Limitations:} Loss of accuracy in some unnatural threshold cases
that do not have vanishing masses

\vspace{0.35cm}

\noindent
{\it Typical running time:} Less than a second

\section{Introduction}
\label{sec:intro}
\setcounter{equation}{0}
\setcounter{footnote}{1}

The precision of measurements within the Standard Model requires
a level of accuracy obtained from two-loop, and even higher-order,
calculations in quantum field theory. The future discoveries of the
Large Hadron Collider and a future $e^+e^-$ linear collider 
may well extend these requirements even further. For example,
supersymmetry predicts the existence of many new particle states
with perturbative interactions. The most important observables in
softly broken supersymmetry are just the new particle masses.
Therefore, comparisons of particular models of supersymmetry
breaking with experiment will require systematic methods
for two-loop computations of physical pole masses in terms of
the underlying Lagrangian parameters.

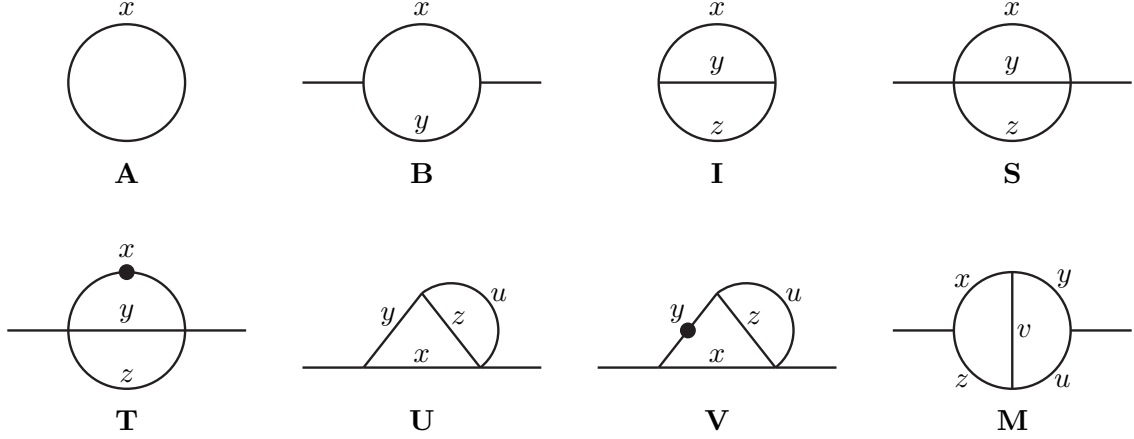
\begin{figure}[t]  
\begin{picture}(108,80)(-54,-40)
\SetWidth{0.9}
\CArc(0,0)(22,0,360)
\Text(0,28)[]{$x$}
\Text(0,-34)[]{$\bf A$}
\end{picture}
\begin{picture}(108,80)(-54,-40)
\SetWidth{0.9}
\Line(-45,0)(-22,0)
\Line(45,0)(22,0)
\CArc(0,0)(22,0,360)
\Text(0,28)[]{$x$}
\Text(0,-16)[]{$y$}
\Text(0,-34)[]{$\bf B$}
\end{picture}
\begin{picture}(108,80)(-54,-40)
\SetWidth{0.9}
\CArc(0,0)(22,0,360)
\Line(-22,0)(22,0)
\Text(0,28)[]{$x$}
\Text(0,6.5)[]{$y$}
\Text(0,-17)[]{$z$}
\Text(0,-34)[]{$\bf I$}
\end{picture}
\begin{picture}(108,80)(-54,-40)
\SetWidth{0.9}
\Line(-45,0)(-22,0)
\Line(45,0)(22,0)
\CArc(0,0)(22,0,360)
\Line(-22,0)(22,0)
\Text(0,28)[]{$x$}
\Text(0,6.5)[]{$y$}
\Text(0,-17)[]{$z$}
\Text(0,-34)[]{$\bf S$}
\end{picture}
\hfill

\vspace{0.45cm}

\begin{picture}(108,80)(-54,-40)
\SetWidth{0.9}
\Line(-45,0)(-22,0)   
\Line(45,0)(22,0)  
\CArc(0,0)(22,0,360)
\Line(-22,0)(22,0)   
\Vertex(0,22){3.0}
\Text(0,30.2)[]{$x$}  
\Text(0,6.5)[]{$y$}
\Text(0,-17)[]{$z$}
\Text(0,-34)[]{$\bf T$}
\end{picture}
\begin{picture}(108,80)(-54,-40)
\SetWidth{0.9}
\Line(-45,-14)(-22,-14)
\Line(45,-14)(22,-14)
\Line(-22,-14)(0,14)
\Line(22,-14)(0,14)
\CArc(11,0)(17.8045,-51.843,128.157)
\Line(-22,-14)(22,-14)
\Text(0,-9.67)[]{$x$}
\Text(-12.8,5.3)[]{$y$}
\Text(13.8,4.8)[]{$z$}
\Text(29,13.6)[]{$u$}
\Text(0,-34)[]{$\bf U$}
\end{picture}
\begin{picture}(108,80)(-54,-40)
\SetWidth{0.9}
\Line(-45,-14)(-22,-14)
\Line(45,-14)(22,-14)
\Line(-22,-14)(0,14)
\Vertex(-11,0){3.0}
\Line(22,-14)(0,14)
\CArc(11,0)(17.8045,-51.843,128.157)
\Line(-22,-14)(22,-14)
\Text(0,-9.67)[]{$x$}
\Text(-15,6.2)[]{$y$}
\Text(13.8,4.8)[]{$z$}
\Text(29,13.6)[]{$u$}
\Text(0,-34)[]{$\bf V$}
\end{picture}
\begin{picture}(108,80)(-54,-40)
\SetWidth{0.9}
\Line(-45,0)(-22,0)
\Line(45,0)(22,0)
\CArc(0,0)(22,0,360) 
\Line(0,-22)(0,22) 
\Text(-18.9,18.9)[]{$x$}
\Text(19.7,19.7)[]{$y$}   
\Text(-19,-19)[]{$z$}
\Text(19,-19)[]{$u$}
\Text(4.5,0)[]{$v$}   
\Text(0,-34)[]{$\bf M$}
\end{picture}
\caption{\label{fig:topologies} 
Feynman diagram topologies for the one- and two-loop vacuum and 
self-energy integrals as defined in this paper.}
\end{figure}

In this paper, we describe a software package called 
{\tt TSIL} (Two-loop Self-energy Integral Library)\footnote{In the Hopi 
culture indigenous
to the American southwest, Tsil is the Chili Pepper Kachina, one of
many supernatural spirits represented by masked doll-like figurines
and impersonated by ceremonial dancers. Tsil is one of the runner
Kachinas. When he overtakes you in a race, he may stuff your mouth
with hot chili peppers.} that can be used to compute two-loop
self-energy functions. In a general quantum field theory, the
necessary dimensionally regularized Feynman integrals can be expressed
in the form:
\beq
\mu^{8-2d} \int d^d k \int d^d q
\frac{
(k^2)^{n_1} (q^2)^{n_2} (k\cdot p)^{n_3} (q \cdot p)^{n_4}
(k \cdot q)^{n_5}
}{
[k^2 +x]^{r_1}
[q^2 + y]^{r_2} 
[(k-p)^2 +z]^{r_3}
[(q-p)^2 +u]^{r_4}
[(k-q)^2 +v]^{r_5}} ,
\label{eq:generaltwoloop}
\eeq
for non-negative integers $n_i,r_i$. (See the master topology $\bf M$ in 
Figure
\ref{fig:topologies}.) The integrations are over Euclidean momenta in
\beq
d = 4 - 2\epsilon 
\eeq
dimensions. An algorithm has been derived by O.V.~Tarasov
\cite{Tarasov:1997kx}, and implemented in a Mathematica
program TARCER by 
Mertig
and Scharf \cite{Mertig:1998vk}, that allows all such integrals to be
systematically reduced to linear combinations of a few basis integrals. 
The coefficients are ratios of polynomials in
the external momentum invariant and the propagator squared masses
$x,y,z,u,v$. In the remainder of this section, we will describe our
notations and 
conventions\footnote{These are the same as in 
refs.~\cite{Martin:2003qz,Martin:2003it};
the correspondences with some other papers is given in Appendix A.}
for the two-loop basis integrals and some related functions, and describe 
the strategy used by {\tt TSIL} to compute them.

First, we define a loop factor
\begin{equation}
C = (16 \pi^2) \frac{\mu^{2\epsilon}}{(2 \pi)^d}
  = (2 \pi \mu)^{2 \epsilon}/\pi^2 .
\end{equation}
The regularization scale $\mu$ is related to the renormalization scale $Q$
(in the modified minimal subtraction renormalization scheme based on
dimensional regularization \cite{DREG}, or dimensional reduction
\cite{DRED} for softly-broken supersymmetric theories) by
\begin{equation}
Q^2 = 4 \pi e^{-\gamma} \mu^2 .
\end{equation}
Logarithms of dimensionful quantities are always given in terms of
\begin{equation}
\lnbar X \equiv {\rm ln}(X/Q^2) .
\end{equation}
The loop integrals are functions of a common external momentum
invariant
\beq
s = -p^2 
\eeq
using a Euclidean or signature ($-$$+$$+$$+$) metric. (Note that the sign
convention is such that for a stable physical particle with mass $m$,
there is a pole at $s = m^2$.) On the physical sheet, $s$ has an
infinitesimal positive imaginary part. Since all functions in any given
equation typically have the same $s$ and $Q^2$, they are not included
explicitly in the list of arguments in written formulas.  A prime on a
squared-mass argument of a function indicates differentiation with respect
to that argument.  Thus, for example
\beq
f(x^\prime,x,z^\prime) = \left[{\partial^2\over \partial y\partial z}
f(y,x,z)\right]_{y=x}\; .
\eeq

We now define one-loop vacuum and self-energy integrals as:
\begin{eqnarray}
{\bf A}(x) &=&  
C \int d^d k \frac{1}{[k^2 +x]}  
\label{defineboldA}
\\
{\bf B}(x,y) &=&
C \int d^d k   \frac{1}{[k^2 +x] [(k-p)^2 +y]} .
\label{defineboldB}
\end{eqnarray}
We also define two-loop integrals according to:
\begin{eqnarray}
{\bf S}(x,y,z) &=&
C^2\int d^d k \int d^d q
\frac{1}{[k^2 +x] [q^2 +y] [(k+q-p)^2 +z]} 
\label{defineboldS}
\\
{\bf I}(x,y,z) &=& {\bf S}(x,y,z) |_{s=0} 
\label{defineboldI}
\\
{\bf T}(x,y,z) &=&
-{\bf S}(x',y,z) 
\label{defineboldT}
\\
{\bf U}(x,y,z,u) &=&
C^2 \int d^d k \int d^d q
\frac{1}{[k^2 +x] [(k-p)^2 +y] [q^2 +z] [(q+k-p)^2 + u]} 
\label{defineboldU}
\\
{\bf V}(x,y,z,u) &=& 
-{\bf U}(x,y',z,u) 
\label{defineboldV}
\\
{\bf M}(x,y,z,u,v) &=& C^2 
\int d^d k \int d^d q
\frac{1}{[k^2 +x][q^2 + y] [(k-p)^2 +z][(q-p)^2 +u][(k-q)^2 +v]} .
\phantom{xxxxx}
\label{defineboldM} 
\end{eqnarray}
The corresponding Feynman diagram topologies are shown in
fig.~\ref{fig:topologies}.
These integrals have various symmetries that are clear from the diagrams:
${\bf B}(x,y)$ is invariant under interchange of $x,y$; the ``sunrise"
integral ${\bf S}(x,y,z)$ and ${\bf I}(x,y,z)$ are invariant under
interchange of any two of $x,y,z$;  ${\bf T}(x,y,z)$ is invariant under 
$y \leftrightarrow z$; ${\bf U}(x,y,z,u)$ and ${\bf V}(x,y,z,u)$ are
invariant under $z \leftrightarrow u$; and the ``master" integral 
${\bf M}(x,y,z,u,v)$ is invariant under each of the interchanges
$(x,z)\leftrightarrow (y,u)$, and $(x,y)\leftrightarrow (z,u)$, and
$(x,y)\leftrightarrow (u,z)$. 

It is often convenient to introduce modified integrals in which 
appropriate divergent parts have been subtracted and the regulator
removed. At one-loop order, 
we define the finite and $\epsilon$-independent integrals:
\begin{eqnarray}
A(x) &=& \lim_{\epsilon \rightarrow 0} \left [{\bf A}(x) + x/\epsilon 
\right ]
= x (\lnbar x-1) 
\\
B(x,y) &=&
\lim_{\epsilon \rightarrow 0} \left [ {\bf B}(x,y) - 1/\epsilon \right ]
= -\int_0^1 dt \>\lnbar [t x + (1-t) y -t (1-t) s ]
.
\end{eqnarray}
At two loops, we let
\begin{eqnarray}
S(x,y,z) &=&  
\lim_{\epsilon \rightarrow 0} \left [
{\bf S}(x,y,z)
- S^{(1)}_{\rm div} (x,y,z) - S^{(2)}_{\rm div} (x,y,z)
\right ] , 
\end{eqnarray}
where
\begin{eqnarray}
S^{(1)}_{\rm div} (x,y,z) &=&
\left ({\bf A}(x) + {\bf A}(y) + {\bf A}(z) \right )/\epsilon 
\\
S^{(2)}_{\rm div} (x,y,z) &=&
(x+y+z)/2\epsilon^2 + (s/2-x-y-z)/2 \epsilon
\end{eqnarray}
are the contributions from
one-loop subdivergences
and from the remaining two-loop divergences, respectively.
In addition,
\begin{eqnarray}
I(x,y,z) &=& S(x,y,z) |_{s=0} 
\\
T(x,y,z) &=& 
-S(x',y,z) .
\end{eqnarray}
Similarly, we define
\begin{eqnarray}
U(x,y,z,u) &=& \lim_{\epsilon \rightarrow 0} \left [
{\bf U}(x,y,z,u)- U^{(1)}_{\rm div} (x,y) - U^{(2)}_{\rm div}  
\right ]
\end{eqnarray}
where
\begin{eqnarray}
U^{(1)}_{\rm div} (x,y) &=& {\bf B}(x,y)/\epsilon 
\\
U^{(2)}_{\rm div} &=& -1/2\epsilon^2 + 1/2\epsilon
\end{eqnarray}
are again the contributions from one-loop sub-divergences and
the remaining two-loop divergences. 
Also, we define
\beq
V(x,y,z,u) &=& 
-U(x,y',z,u) .
\eeq
The master integral is free of divergences, so we define
\begin{eqnarray}
M(x,y,z,u,v) &=&
\lim_{\epsilon \rightarrow 0} {\bf M}(x,y,z,u,v) .
\end{eqnarray}   

Thus, bold-faced letters $\bf A,B,I,S,T,U,V$ represent the original
regularized integrals that diverge as $\epsilon \rightarrow 0$, while the
ordinary letters $A,B,I,S,T,U,V,M$ are used to represent functions 
that are finite and independent of $\epsilon$ by definition.
Note, however, that as defined above $I,S,T,U,V$ are not simply the 
$\epsilon$-independent terms in expansions in small $\epsilon$.
The following expansions are useful for converting between
$\bf I,S,T,U,V$ and $I,S,T,U,V$:
\beq
{\bf A}(x) &=& -x/\epsilon + A(x) + \epsilon A_\epsilon(x) + 
{\cal O}(\epsilon^2) 
\\
{\bf B}(x,y) &=& 1/\epsilon + B(x,y) + \epsilon B_\epsilon(x,y) 
+ {\cal O}(\epsilon^2) ,
\eeq
where
\beq
A_\epsilon(x) &=& x[ -1 - \zeta (2)/2 + \lnbar x - (\lnbar x)^2/2] 
\\
B_\epsilon(x,y) &=& 
\zeta (2)/2 + \frac{1}{2} 
\int_0^1 dt \> (\lnbar [t x + (1-t) y -t (1-t) s ])^2 .
\eeq
Here $\zeta$ is the Riemann zeta function.
The function $B_\epsilon(x,y)$ can be expressed in terms of dilogarithms
\cite{Nierste:1992wg},
and is given by the coefficient of $\delta$ in eq.~(83) of 
ref.~\cite{Scharf:1993ds}. Also, 
\beq
{\bf B}(x',y) = 
\Bigl [ (3-d) (s-x+y) {\bf B}(x,y) + (2-d) \lbrace 
{\bf A}(y) 
+ (s-x-y){\bf A}(x)/2x \rbrace \Bigr ]/\Delta_{sxy}
\eeq
where
\beq
\Delta_{abc} \equiv a^2 + b^2 + c^2 - 2 a b - 2 a c - 2 b c.
\eeq
{}From the preceding equations it follows that
\beq
{\bf I}(x,y,z) &=& 
-(x+y+z)/2\epsilon^2
+ \left[ A(x) + A(y) + A(z) - (x+y+z)/2 \right ]/\epsilon
\\
\nonumber 
&&
+ I(x,y,z) + A_\epsilon (x) + A_\epsilon (y) + A_\epsilon (z)
+ {\cal O}(\epsilon)
\label{eq:Iboldformula}
\\
{\bf S}(x,y,z) &=& 
-(x+y+z)/2\epsilon^2
+ \left[ A(x) + A(y) + A(z) - (x+y+z)/2 +s/4 \right ]/\epsilon
\\
\nonumber 
&&
+ S(x,y,z) + A_\epsilon (x) + A_\epsilon (y) + A_\epsilon (z)
+ {\cal O}(\epsilon)
\\
{\bf T}(x,y,z) &=& 
1/2\epsilon^2 
- \left[ A(x)/x + 1/2 \right ]/\epsilon
+ T(x,y,z) 
+ [A(x) - A_{\epsilon}(x)]/x
+ {\cal O}(\epsilon) \phantom{xxxx}
\label{eq:bfTexp}
\\
{\bf U}(x,y,z,u) &=& 
1/2\epsilon^2 + \left [ 1/2 + B(x,y) \right ]/\epsilon 
+ U(x,y,z,u) + B_\epsilon (x,y)
+ {\cal O}(\epsilon)
\\
{\bf V}(x,y,z,u) &=& 
\frac{1}{\epsilon}
[(s+x-y)(B(x,y) - 1) + 2 A(x) + (s-x-y) 
A(y)/y]/\Delta_{sxy}  \\
\nonumber 
&&
 + V(x,y,z,u)
 +\Bigl \lbrace (s+x-y) [B_{\epsilon}(x,y) -2B(x,y)]
+ 2 A_{\epsilon}(x) -2 A(x) 
\\
\nonumber 
&&
+ (s-x-y)[A_{\epsilon}(y) - A(y)]/y \Bigr \rbrace/\Delta_{sxy}
+ {\cal O}(\epsilon)
\\
{\bf M}(x,y,z,u,v) &=& M(x,y,z,u,v)
+ {\cal O}(\epsilon) .
\label{eq:Mboldformula}
\eeq

The internal workings of the {\tt TSIL} code use the
functions $A,B,I,S,T,U,V,M$ rather than their bold-faced counterparts.
This is because we find that renormalized expressions for physical
quantities are more compactly written in terms of the non-bold-faced
integrals. However, both types of functions are available as outputs, with
the proviso that for $\bf I,S,T,U,V,M$ we keep only the pole and
finite terms as $\epsilon \rightarrow 0$, and for $\bf A,B$ only
terms through order $\epsilon$. 

Tarasov's algorithm \cite{Tarasov:1997kx} allows any integral of the form
of eq.~(\ref{eq:generaltwoloop}) to be expressed\footnote{Actually,
Tarasov's algorithm relates the general integral to the bold-faced
versions of the basis functions, and holds for general $d$. 
By neglecting contributions that vanish 
for $\epsilon \rightarrow 0$, 
one can write the results 
in terms of the non-bold-faced functions.} 
as a linear combination of the
following two-loop basis integrals: 
\beq
&&
{M}(x,y,z,u,v),
\>\>\>
{U}(z,x,y,v),
\>\>\>
{U}(u,y,x,v),
\>\>\>
{U}(x,z,u,v),
\>\>\>
{U}(y,u,z,v),
\>\>\>
{T}(v,y,z),
\phantom{xxxxx}
\nonumber
\\
&&
{T}(u,x,v),
\>\>\>
{T}(y,z,v), 
\>\>\>
{T}(x,u,v),
\>\>\>
{T}(z,y,v),
\>\>\>
{T}(v,x,u),
\>\>\>
{S}(v,y,z), 
\>\>\>
{S}(u,x,v),
\label{eq:basis}
\eeq
plus terms involving the two-loop vacuum integrals $I(x,y,v)$ or $I(z,u,v)$,
or quadratic in the one-loop integrals.  

In particular, the $V$ and ${\bf V}$ integrals can be expressed as linear 
combinations
of the others (see eqs.~(3.22)--(3.28) and (6.21) 
of ref.~\cite{Martin:2003qz}), and so 
are not included in the basis.  However, they are included as outputs,
because some results are more compactly written in terms of them.
Because $T(x,y,z)$ is divergent in the limit $x\rightarrow 0$, 
it is also sometimes useful to define the function:
\beq
{\overline T}(x,y,z) = T(x,y,z) + \,B(y,z)\,\lnbar x .
\eeq
For $x=0$, 
this function is well-defined and can be written in terms of 
dilogarithms (see 
eqs.~(6.18)--(6.19) of ref.~\cite{Martin:2003qz}).
In that limit it can also be rewritten in terms of the other basis 
functions, see eqs.~(A.15)--(A.16) of
ref.~\cite{Martin:2003it}, but is still available as an output
of {\tt TSIL} for convenience. A useful identity supplementing
eq.~(\ref{eq:bfTexp}) is:
\beq
{\bf T}(0,y,z) &=& 
-1/2\epsilon^2 
+ [1/2 - B(y,z)]/\epsilon
+ {\overline T}(0,y,z) 
- B_{\epsilon}(y,z)
+ {\cal O}(\epsilon).
\eeq

It remains to provide a means for the numerical computation
of the basis integrals. For special values of masses and external
momentum, it is possible to compute the two-loop integrals
analytically in terms of polylogarithms~\cite{Lewin}. This requires
\cite{Scharfthesis} that there is no three-particle cut of the diagram for
which the cut propagator masses $m_1$, $m_2$, $m_3$ and the five 
quantities
\begin{equation} 
s_{\rm cut}, \>\>\> s_{\rm cut} - (m_1\pm m_2 \pm m_3)^2 ,
\end{equation} 
(where $s_{\rm cut} = -p^2_{\rm cut}$ is the momentum invariant for 
the total momentum flowing 
across the 
cut) are all non-zero. Many analytical results for various such special 
cases have been worked out in refs.~\cite{Rosner}--\cite{Laporta:2004rb}, 
\cite{Scharf:1993ds}, \cite{Martin:2003qz}, and 
Appendix B of the present paper. 
There 
are also
expansions \cite{Smirnov:rz}--\cite{Berends:1994sa} in large and small
values of the external momentum invariant, and near the thresholds and
pseudo-thresholds \cite{Berends:1996gs}--\cite{Jegerlehner:2002em}. 
Analytical results in terms of polylogarithms for the 
$A,B,I,S,T,U,M$ functions at generic
values of $s$ are reviewed in
section VI of ref.~\cite{Martin:2003qz}. These and some other 
analytical formulas
for special cases
have been implemented in the {\tt TSIL} code. They consist of
the master integral cases:
\beq
\renewcommand{\arraystretch}{1.4}
\begin{array}[t]{ll}
M(x,x,y,y,0),
\>\>\> 
\>\>\> 
M(0,0,0,x,0), 
\>\>\> 
\>\>\> 
M(0,x,0,x,x), 
\phantom{xxxx}
&
\mbox{ref.~\cite{Broadhurst:1987ei}}
\\
M(0,0,0,0,x), 
\>\>\> 
\>\>\> 
M(0,x,0,y,0),
&
\mbox{ref.~\cite{Scharf:1993ds}}
\\
M(0,0,0,x,x), 
\>\>\> 
\>\>\> 
M(0,x,x,0,0), 
\>\>\> 
\>\>\> 
M(0,x,x,x,0), 
&
\mbox{ref.~\cite{Fleischer:1998nb}}
\\
M(0,0,x,y,0), 
&
\mbox{Appendix B}\phantom{xxxxxxx}
\\
M(0,x,x,0,x)|_{s=x}, 
&
\mbox{ref.~\cite{Broadhurst:1987ei}}
\\
M(0,y,y,0,x)|_{s=x}, 
&
\mbox{Appendix B}
\\
M(0,x,y,0,y)|_{s=x}, 
&
\mbox{refs.~\cite{Fleischer:1998dw,Jegerlehner:2003py} (see also Appendix B)}
\end{array}
\nonumber
\eeq
and functions obtained by permutations of the arguments,
and all subordinate integrals
$A$, $B$, $I$, $S$, $T$, $U$, $V$, $\overline{T}$ that
have 
topologies obtained by removing one or more propagator lines from the 
cases above. These include:
\beq
\renewcommand{\arraystretch}{1.4}
\begin{array}[t]{ll}
U(x,y,0,y), \qquad& 
\mbox{ref.~\cite{Djouadi:1987di}}
\\
U(x,y,0,0), 
\>\>\>
U(0,0,0,x),
\qquad&
\mbox{ref.~\cite{Scharf:1993ds}}
\\
U(0,x,y,z), \qquad&
\mbox{ref.~\cite{Martin:2003qz}}
\\
U(x,0,0,y), \qquad&
\mbox{ref.~\cite{Martin:2003it}}
\\
U(x,0,y,y)|_{s=x},
\>\>\>
U(y,0,y,x)|_{s=x}, \qquad&
\mbox{refs.~\cite{Gray:1990yh,Davydychev:1998si} (see also Appendix B)}
\end{array}
\renewcommand{\arraystretch}{1.0}
\nonumber
\eeq
and 
\beq
\renewcommand{\arraystretch}{1.4}
\begin{array}[t]{ll}
S(0,x,y), \>\>\> T(x,0,y),\>\>\> \overline{T}(0,x,y),
\qquad&
\mbox{ref.~\cite{Berends:1994ed}}
\\
S(x,y,y)|_{s=x}, \>\>\> T(x,y,y)|_{s=x}, \>\>\> T(y,x,y)|_{s=x}. 
\qquad& 
\mbox{refs.~\cite{Gray:1990yh,Berends:1997vk}}
\end{array}
\renewcommand{\arraystretch}{1.0}
\nonumber
\eeq
Also included are all of the functions at $s=0$, which can be easily
expressed in terms of the $A$ and $I$ functions and derivatives
of them, which are in turn known 
\cite{vanderBij:1983bw,Ford:hw}
analytically in terms of logarithms
and dilogarithms.

For the case of generic masses and $s$, another method is needed.
Integral representations have been studied in 
refs.~\cite{Kreimer:1991jv}--\cite{Passarino:2001jd}.
For {\tt TSIL}, we instead use the differential
equations method \cite{Kotikov:1990kg}--\cite{Remiddi:1997ny} to evaluate 
the
integrals numerically.  For the ${\bf S}$, ${\bf T}$ and ${\bf U}$
integrals, this strategy was proposed and implemented in 
\cite{Caffo:1998du}--\cite{Caffo:2003ma}. The method
was rewritten in terms of the $S,T,U$ integrals and extended to $M$ in
ref.~\cite{Martin:2003qz}. To see how the method works,
consider the functions listed in
eq.~(\ref{eq:basis}) and also $B(x,z)$ and $B(y,u)$ and the product
$B(x,z) B(y,u)$. Let us denote these sixteen quantities by 
$F_i$ where $i = 1,\ldots,16$. Considered
as functions of $s$ for fixed $x$, $y$, $z$, $u$, $v$, $Q^2$
they 
can be shown to satisfy a set 
of coupled first-order differential equations of the form
\beq
\frac{d}{ds}F_i = \sum_{j} C_{ij} F_j + C_i ,
\label{eq:genericdiffeq}
\eeq
where the coefficients $C_{ij}$ and $C_i$ are ratios of polynomials in the
squared masses and $s$ (and of the analytically known $A$ and $I$
functions, for some of the coefficients not multiplying two-loop 
functions). 
The coefficients can be evaluated by using Tarasov's recurrence
relations, and were listed in \cite{Martin:2003qz}.

\begin{figure}[t]  
\begin{center}
\begin{picture}(210,110)(-30,-30)
\SetScale{0.5}
\SetWidth{0.9}
\ZigZag(150,0)(385,0){3}{20}
\SetWidth{1.2}
\Vertex(150,0){3.5}
\Vertex(210,0){3.5}
\Vertex(240,0){3.5}
\Vertex(325,0){3.5}
\LongArrow(150,-35)(150,-10)   
\LongArrow(210,-35)(210,-10) 
\LongArrow(240,-35)(240,-10) 
\LongArrow(325,-35)(325,-10) 
\LongArrow(-30,0)(395,0) 
\LongArrow(0,-30)(0,195)   
\SetWidth{3.0}
\ArrowLine(0,0)(0,140) 
\ArrowLine(0,140)(305,140)
\ArrowLine(305,140)(305,4)
\SetPFont{Times-Roman}{24}
\PText(380,-10)(0)[]{Re[s]}
\PText(-34,190)(0)[]{Im[s]}
\PText(240,-50)(0)[]{thresholds}
\end{picture}
\end{center}
\caption{\label{fig:contour} 
When $s$ is greater than or equal to the smallest non-zero threshold or 
pseudo-threshold, the
Runge-Kutta integration proceeds along a contour with the shape shown
here, as suggested in ref.~\cite{Caffo:2002wm}.} \end{figure}
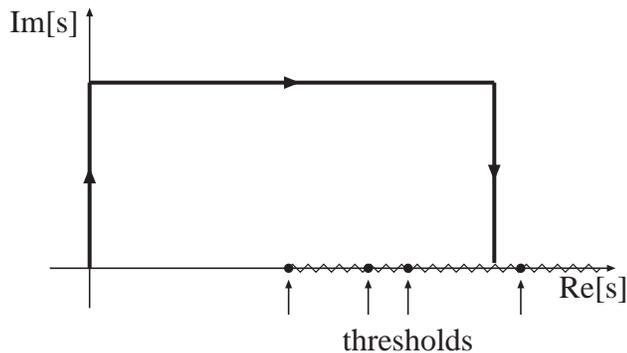

At $s=0$ the values of all of the functions and their 
derivatives with respect to $s$ (and/or their expansions
in small $s$) are known analytically in terms of dilogarithms. Therefore, 
one can integrate the functions\footnote{For the master
integral, we actually run $sM(x,y,z,u,v)$.} by the Runge-Kutta method to
any other value of $s$. In order to avoid numerical problems from
integrating through thresholds and pseudo-thresholds, we use the
suggestion of ref.~\cite{Caffo:2002wm} by following a displaced contour in
the upper-half complex $s$ plane, as shown in fig.~\ref{fig:contour},
whenever $s$ is greater than the smallest non-zero threshold or
pseudo-threshold.  This contour starts from $s=0$ (or, in some cases, a
point very close by, as explained in the next section) and proceeds to a
point $is_{\rm im}$, from there to $s + i s_{\rm im}$, and from there to
the desired value $s$.  Here $s_{\rm im}$ is real and positive. 
Since $s$
has an infinitesimal positive imaginary part on the physical sheet, this
procedure also automatically produces the correct branch cut behavior for
functions when $s$ is above thresholds.  When $s$ is less than or equal to
the smallest non-zero threshold or pseudo-threshold, the Runge-Kutta
integration instead proceeds directly along the real $s$ axis. In typical 
physical problems, if the master integral is needed, then so will be all 
of its subordinate $B,S,T,U$ integrals. These are all obtained 
simultaneously as a result of the Runge-Kutta method. Furthermore, checks 
on the numerical accuracy can be made by varying the Runge-Kutta step size 
parameters
and the choice of contour in the upper-half complex $s$ plane. 

Most of the practical difficulties in realizing this program have to do
with numerical instabilities when the final value of $s$ is at or near a
threshold or pseudo-threshold, or when the starting point $s=0$ is itself
a threshold. We describe these issues and the methods used by 
{\tt TSIL} to successfully evade them in the next section.

\section{Numerical considerations near thresholds and pseudo-thresholds}
\label{sec:thresholds}
\setcounter{equation}{0}
\setcounter{footnote}{1}

The two-loop master integral function $M(x,y,z,u,v)$, and its subordinates
listed in eq.~(\ref{eq:basis}), have two-particle and three-particle
thresholds at $s$ equal to:
\beq
&& t_{xz} = (\sqrt{x} + \sqrt{z})^2 ,
\qquad
\qquad\quad\>\>\>\>
t_{yu} = (\sqrt{y} + \sqrt{u})^2 ,
\\
&& 
t_{xuv} = (\sqrt{x} + \sqrt{u} + \sqrt{v})^2 ,
\qquad\quad
t_{yzv} = (\sqrt{y} + \sqrt{z} + \sqrt{v})^2 .
\eeq
At the thresholds, the integral functions have branch cuts, and
they therefore develop imaginary part contributions for $s > s_{\rm 
thresh}$. 
The pseudo-thresholds occur at $s$ equal to:
\beq
&&
p_{xz} = (\sqrt{x} - \sqrt{z})^2 ,
\qquad\qquad\quad\>\>\>\>\>\>
p_{yu} = (\sqrt{y} - \sqrt{u})^2 ,
\\
&&
p_{xuv} = (-\sqrt{x} + \sqrt{u} + \sqrt{v})^2 ,
\qquad\quad
p_{uxv} = (\sqrt{x} - \sqrt{u} + \sqrt{v})^2 ,
\\
&&
p_{vxu} = (\sqrt{x} + \sqrt{u} - \sqrt{v})^2 ,
\qquad\quad\>\>\>
p_{yzv} = (-\sqrt{y} + \sqrt{z} + \sqrt{v})^2 ,
\\
&&
p_{zyv} = (\sqrt{y} - \sqrt{z} + \sqrt{v})^2 ,
\qquad\quad\>\>\>
p_{vyz} = (\sqrt{y} + \sqrt{z} - \sqrt{v})^2 .
\eeq
The integral functions are analytic at the pseudo-threshold points
(unless they coincide with a threshold), but the coefficients in
the differential equation have pole singularities at both
the thresholds and pseudo-thresholds. For values of
$s$ close to both types of special points, one must therefore be
careful in numerical evaluation to avoid undefined quantities and
round-off errors. In this 
section, we 
discuss the ways in which {\tt TSIL} avoids these problems.

First, we consider the case that the initial point of the 
Runge-Kutta 
integration, $s=0$, is actually a threshold.
This occurs for master integrals $M(0,y,z,0,0)$ and $M(0,y,0,u,v)$,
and permutations thereof. In these cases, some of the basis integral
functions have logarithmic singularities and/or branch cuts at $s=0$. To 
deal with this, we
make a change of independent variable to
\beq
r = \lnbar (-s),
\label{seq0cov}
\eeq
and start the
integration at a point $r_0 = \ln(-i \delta) = 
\ln(\delta) - i\pi/2$, with $\delta$ real, positive, and
extremely small (of order the relative error of the computer 
arithmetic). The initial values of the integrals 
are obtained from the expansions in small $s$, given in section V of 
ref.~\cite{Martin:2003qz}. This change of variables is also
used when $s=0$ is close to, but not exactly a threshold (with
the exact criteria adjustable by the user). This variable $r$ is
used for the first leg of the contour in the Runge-Kutta integration.

Next, consider the case that the final value of $s$ is at, or very near,
a threshold $s_{\rm thresh}$. In this case, we find that the change of 
variable
\beq
r = \ln (1 - s/s_{\rm thresh})
\label{sfinalcov}
\eeq
is effective, and so is used by {\tt TSIL} for the final part of
the Runge-Kutta integration. 
When the final value of $s$ is exactly equal to a threshold,
then the endpoint of the running is taken to be $r = \ln(\delta) - i\pi/2$, 
with $\delta$ again taken to be real, positive, and extremely small. 

We next describe the Runge-Kutta algorithms used, since they have some
slightly unusual properties dictated by the need for control of precision
near thresholds and pseudo-thresholds. 
Consider a vector of quantities $\vec F$ that satisfy coupled first-order 
differential equations $d\vec F/dt = \vec f(t,\vec F)$.
The general form for an explicit $m$-stage Runge-Kutta routine advancing the
solution from $t$ to $t + h$ is given by:
\beq
\vec F(t+h) = \vec F(t) + h \sum_{i=1}^m b_i \vec k_i,
\label{RKstep}
\eeq
where
\beq
\vec k_i = \vec f (t + c_i h, \vec F(t) + h \sum_{j=1}^{i-1} a_{ij} \vec 
k_j) \qquad \mbox{(for $i=1,\ldots, m$)}.
\eeq
Here $a_{ij}$, $b_i$, and $c_i$ 
are known as Butcher coefficients, and 
satisfy $c_1 = 0$, $a_{10} = 0$, and 
\beq
c_i  &=& \sum_{j=1}^{i-1} a_{ij} \qquad\mbox{(for $i=2,\ldots, m$)}
\\
\sum_{i=1}^m b_i &=& 1
\eeq
plus other, non-linear, constraints \cite{Butcher}.
The algorithm is said to be of $n$th order if the error
is proportional to $h^{n+1}$ for sufficiently small $h$.
In order to implement automatic step-size adjustment, we use a
6-stage embedded Runge-Kutta \cite{Fehlberg} with
coefficients given by Cash and Karp \cite{CashKarp}. 
These give not only a 5th-order step as in eq.~(\ref{RKstep}), but a
4th-order step estimate of the same form with different coefficients $b_i^*$.
This gives an error estimate for each dependent variable, for each step:
\beq
\Delta \vec F(t+h) = h \sum_{i=1}^m (b_i - b_i^*) \vec k_i.
\label{RKstepembedded}
\eeq
The theoretical estimate of the step size needed so that the error
for each dependent variable
is less than $\delta_P$ times the increment of that variable is
then given by:
\beq
h_{\rm new} = h S
\left [ \frac{\delta_P}{{\rm Max}(|\Delta\vec F|)} \right ]^{1/4}
,
\label{eq:newhest}
\eeq
where $S$ is a safety factor less than unity.

However,
in the present application
there is a special problem because the final destination point 
might be equal to (or close to) a 
threshold or pseudo-threshold. There, the individual coefficients in the 
derivatives of the functions might be ill-defined (or subject to large
numerical round-off errors, because of small denominators in individual
terms), even though the basis functions themselves are well-defined. 
To avoid this problem, we instead need to use an $m$-stage
Runge-Kutta with the crucial property $c_i < 1$ for all $i$, so that no 
derivatives are ever evaluated
at the endpoint. There are no 4-stage, 4th-order
solutions to the Butcher coefficient conditions with this property,
but there are many 5-stage, 4th-order solutions. We chose,
somewhat arbitrarily, the set:
\beq
&&c_i = (0,\,\> 1/4,\,\> 3/8,\,\> 1/2,\,\> 5/8)
\label{eq:specialRKstart}
\\
&&b_i = (-1/15,\,\> 2/3,\,\> 4/3,\,\> -10/3,\,\> 12/5) \\
&&
a_{21} = 1/4,
\> \,
a_{32} = 3/8,
\> \,
a_{42} = 1/2,
\> \,
a_{52} = 35/72,
\> \,
a_{54} = 5/36,\phantom{xxx}
\\
&&a_{ij} = 0\qquad\mbox{for other $i,j$} .
\label{eq:specialRKend}
\eeq
This procedure is not as efficient as the Cash-Karp 6-stage, 5th-order
algorithm under normal circumstances, and does not provide 
an error estimate,
so it is used only where needed for the very last Runge-Kutta step.

The {\tt TSIL} implementation of the coefficients $C_{ij}$ and $C_i$
in eqs.~(\ref{eq:genericdiffeq}) is also done in a special way to avoid
roundoff errors near thresholds and pseudo-thresholds. The expressions as
given in \cite{Martin:2003qz} for these coefficients appear to have double
(or higher) poles in $s$ for some special values of the masses with
degenerate thresholds and pseudo-thresholds. The presence of such
higher-order poles can lead to large round-off errors, due to
incomplete cancellations in computer arithmetic with finite precision.
Fortunately, this is avoided in most cases by applying the partial
fractions technique to rewrite the coefficients in the derivatives with
respect to $s$, so that all poles in $s$ are at most simple poles. This
can always be done for the $B$, $S$, and $T$ functions.

For the $U$ functions, double poles in the coefficient functions
for the derivative with respect to $s$ remain only if
the second argument vanishes. Here, we take advantage of the
facts that 
$U(x,0,y,z)$ does not enter into the differential
equations that govern the other basis functions, and it
can always be written algebraically in terms of them 
(see eqs.~(A.15), (A.17) and (A.18) of
ref.~\cite{Martin:2003it}). Therefore, when the second argument of a
$U$ function vanishes, the result obtained for it from the Runge-Kutta
running is irrelevant; it is simply replaced by the algebraic result 
before it is returned
by the program, eliminating the roundoff error problem.

In most cases for which double poles in the coefficient functions of the
derivative of the master integral cannot be eliminated, it 
can be evaluated
in terms of polylogarithms, so the Runge-Kutta technique is not needed
anyway. A special case in which this does not occur
is $M(x,y,z,u,v)$ for $v = (\sqrt{x} + 
\sqrt{y})^2$;  then $d(sM)/ds$ in ref.~\cite{Martin:2003qz} has 
coefficients with
double poles at $s_0 = [\sqrt{x} (u-y) + \sqrt{y} (z-x)]/\sqrt{v}$.
However, here we can use the identity:
\beq
0 &=& 
\sqrt{x} [U(z,x,y,v) - T(x,u,v)] + \sqrt{y} [U(u,y,x,v) - T(y,z,v)] 
\nonumber \\ &&
+ \sqrt{v} [T(v,u,x) + T(v,y,z) - 1] 
+ [A(v)/\sqrt{v} - A(x)/\sqrt{x} - \sqrt{y}] B(y,u)
\nonumber \\ &&
+ [A(v)/\sqrt{v} - A(y)/\sqrt{y} - \sqrt{x}] B(x,z) 
,
\label{specialUidentity}
\eeq
valid in general for $v = (\sqrt{x} + \sqrt{y})^2$.
When the right side of equation (\ref{specialUidentity}) is multiplied by 
$[\sqrt{z} (u-y) + \sqrt{u} (z-x)]/v (s-s_0)^2$ and added to the 
expression
for $d(sM)/ds$ from ref.~\cite{Martin:2003qz}, the result
generically has no double poles in $s$. That is the form used by the 
program in this special case (and others related by permutations of the
masses).

A non-generic sub-case of the preceding, for which double poles in the 
coefficient
functions of $d(sM)/ds$ are not eliminated, is $M(0,y,z,u,y)$ with $y=
(\sqrt{z} \pm \sqrt{u})^2$. Because the coefficients involve double poles
at $s=z$, there is some loss of accuracy at, and very near, that
threshold.  Fortunately, there is no good reason why a relation like $y=
(\sqrt{z} \pm \sqrt{u})^2$ should hold exactly in a quantum field theory,
unless a symmetry makes $y$ or $z$ or $u$ equal to 0, and in each of these
cases the master integral and all of its subordinates are 
given in terms of polylogarithms. More generally, we have checked that the
coefficients in the derivatives as implemented in {\tt TSIL} always
have only simple poles, except in ``unnatural" cases of this type (where
no symmetry of a quantum field theory can cause the necessary
coincidence), or when the integrals are already analytically computed.

The measures detailed above are generally sufficient to give 
good numerical accuracy
near thresholds and pseudo-thresholds (except in the unnatural coincidence
case just mentioned) without need for interpolation techniques or special
expansions.

\section{Description of the program}
\label{sec:description}
\setcounter{equation}{0}
\setcounter{footnote}{1}

{\tt TSIL} is a library of functions written in C, which can be
(statically) linked from C, C++, or Fortran code.\footnote{A wrapper
routine is included that provides the interface to Fortran; see
section \ref{sec:fortran} below.}  
In addition to the main evaluation
functions, it contains a variety of routines for I/O and other
utilities.  A complete list of functions in the user API is given in
Appendix C.

The principal data object in the code is a C {\tt struct} with type
name {\tt TSIL\_DATA} that contains values of the parameters $x,y,z,u,v$
and $Q^2$ as well as the 15 basis functions of types
$B$, $S$, $T$, $U$, $M$.  Each integral function is
itself a {\tt struct} containing its value, arguments, and various
unchanging coefficients used in computing its derivative.  These
coefficients are functions of $x,y,z,u,v$ and are computed when the
parameter values are set.  In addition, each basis function contains a
set of pointers to the other functions needed in evaluating its
derivative.  Also contained in the data struct are values of
the integrals $\overline T$, $V$, and ${\bf S}$, ${\bf T}$, ${\bf U}$,
and ${\bf V}$. Definitions of all datatypes are contained in the header file
{\tt tsil.h}, which must be included in all application programs.

In any program that calls {\tt TSIL} functions requiring Runge-Kutta 
evaluation, at least one of
these high-level data objects must be declared:
$$
{\tt TSIL\_DATA~~foo;}
$$
(More than one such object, and arrays of such objects, are allowed.
See subsection \ref{subsec:samples} for an example.)  
Users can of course access
the items in the {\tt struct} directly, though it is recommended that
the provided user interface routines be used.  These allow one to
extract values of individual functions (or all of them), set the
values of the external parameters, and so on.

The size of the basic datatypes used for floating point values is
controlled by the user (at compile time) with the choice of compiler flag
{\tt -DTSIL\_SIZE\_LONG} or {\tt -DTSIL\_SIZE\_DOUBLE}.
Then the 
type {\tt TSIL\_REAL} is accordingly synonymous with 
{\tt long double} or {\tt double}, while the 
type {\tt TSIL\_COMPLEX}
is synonymous with 
{\tt long double complex} or {\tt double complex}.
The recommended default size is {\tt long double} on systems 
where it is available.
For most physics applications (taking into account the natural suppression
of two-loop effects), {\tt double} should easily give sufficient accuracy.
However, the use of {\tt long double} provides a nice safety margin,
and execution times are typically short (of order tenths or
hundredths of a
second on a modern workstation for generic inputs) in any case.
Generally, {\tt long double} data 
(typically with 63 or more bits of relative precision) gives results with 
relative accuracies
better than $10^{-10}$ for generic cases, but sometimes somewhat
worse in cases with large mass hierarchies, and in some 
particularly difficult cases 
significantly worse.
[The function $V(x,y,z,u)$ for very small but 
non-zero $y$ can be particularly sensitive
to roundoff errors, since the individual terms in its evaluation 
are proportional to
$1/y$ and yet it is only logarithmically divergent as $y\rightarrow 0$.]
The user should consider modifying the default parameters of the program
if significant sensitivity to parameters is expected (or observed),
or if speed is an overriding concern.

In a typical application the user will initialize the data object and
set values for the external parameters $x,y,z,u,v,Q^2$ by calling the
function {\tt TSIL\_SetParameters}.  (Parameter values may be changed in a
data object with subsequent calls to this function.)  Then the basis
integrals are evaluated at any desired value of $s$ by calling the
master evaluation function {\tt TSIL\_Evaluate}.  Additional calls to
{\tt TSIL\_Evaluate} can be used to 
re-compute the basis functions for other values
of $s$. 

{\tt TSIL\_Evaluate} first decides whether the case at hand is known
analytically; if so, the basis functions are computed directly.  If
not, numerical integration is required.  In this case {\tt TSIL\_Evaluate}
begins by rescaling all dimensionful quantities by the largest of
$x,y,z,u,v,|s|$, 
so that all parameters are rendered dimensionless.  It
then locates all thresholds and pseudo-thresholds and decides whether
special handling is needed, that is, 
if one of these special points is near $s=0$ or the final value of $s$.
The nearness criterion is controlled by a constant {\tt THRESH\_CUTOFF},
defined in {\tt tsil\_params.h}.
If $s=0$ is a threshold (or there is a threshold very near $s=0$),
evaluation proceeds as described above by making the change of
integration variable (\ref{seq0cov}).  If the final value of $s$ is at
or near a threshold, the change of variable (\ref{sfinalcov}) is
enabled for the final leg of the integration contour.

Next, {\tt TSIL\_Evaluate} checks to see if the final value of $s$ is
smaller than the smallest non-zero threshold or pseudo-threshold; if
so, then integration proceeds directly along the real $s$ axis.  If
not, the generic displaced integration contour
(fig.~\ref{fig:contour}) is used.  In cases where the final destination
$s$ is near a threshold or pseudo-threshold, the 
5-stage Runge-Kutta
routine described by eqs.~(\ref{eq:specialRKstart})--(\ref{eq:specialRKend}) 
is used for the very last step of the integration, to avoid
evaluation of any derivatives at the endpoint.

After the Runge-Kutta integration,
the program checks to see if this was a case with
uncanceled double pole terms in the derivatives of the $U$ functions,
due to the second argument vanishing.
If so, these values are corrected.  (Recall that in such cases the
incorrect values obtained by the Runge-Kutta integration have no
effect on other basis functions.)  
The program also replaces any subordinate integrals ($B$, $S$, $T$, $U$) 
that can be computed in terms of polylogarithms by their analytical values.
Finally, the program computes the
values of all $\overline{T}$ and $V$ functions, and the ``bold''
variants of all functions defined in section 2, using 
eqs.~(\ref{eq:Iboldformula})--(\ref{eq:Mboldformula}).

The function {\tt TSIL\_Evaluate} returns 1 ({\tt TRUE}) for successful
execution or 0 ({\tt FALSE}) for error execution.  A warning message
is printed if the external parameters correspond to the unnatural
threshold case discussed at the end of section 3.  The data object
further contains a status parameter, accessible via the function 
{\tt TSIL\_GetStatus}, which indicates how the 
master integral
evaluation was performed:
either analytic, numerical integration along real axis, or numerical
integration along the contour of fig.~\ref{fig:contour}.

The standard output function is {\tt TSIL\_PrintData}, which prints all
function values on {\tt stdout}.  An alternate format, designed so
that captured output can serve as valid input files for Mathematica,
is given by {\tt TSIL\_PrintDataM}.  Additional utilities allow the user to
extract individual basis functions or sets of functions to arrays.
Note that warning and error messages appear on {\tt stderr} so they
may be redirected by the shell and examined separately.  

Along with the size of intrinsic datatypes, the parameters associated
with the numerical integration adaptive step-size control
exert the main influences on execution
speed and accuracy.\footnote{These are always set,
by {\tt TSIL\_SetParameters}, to be equal to the values specified
at compile time in the file {\tt tsil\_params.h}. However, to 
deal with exceptional situations, they
can optionally be reset at run time with 
the function {\tt TSIL\_ResetStepSizeParams},
after calling {\tt TSIL\_SetParameters} and
before calling {\tt TSIL\_Evaluate}.}
They are realized as members of
the {\tt TSIL\_DATA} struct, with names:
\begin{itemize}
\item
{\tt precisionGoal}: This is $\delta_P$ in eq.~(\ref{eq:newhest}). 
(We use a safety
factor $S = 0.9$.) 
If the maximum estimated error for any dependent 
variable exceeds $\delta_P$ multiplied by the increment of 
that variable for that step, and 
also exceeds the relative precision of the
computer arithmetic times the absolute value of that variable, then
the step is retried with a smaller step size, unless the 
step size would become smaller than specified below. Also, after
a successful step, the size for the next step is chosen according to
eq.~(\ref{eq:newhest}), unless it would exceed the amount specified below.
(Defaults: $10^{-12}$ for {\tt long double}, 
$5\times 10^{-11}$ for {\tt double}.)
\item
{\tt nStepsStart}: For each leg of the contour, the initial
step size is chosen so that there would be this many steps if the
step size did not change. (Default: 500)
\item
{\tt nStepsMin}: 
The maximum allowed step size on a leg of the
contour with dimensionless (rescaled) independent variable 
length {\tt L} is given by
{\tt L/nStepsMin}. (Default: 100)
\item
{\tt nStepsMaxCon},
{\tt nStepsMaxVar}: The minimum allowed step size on a leg of the
contour with dimensionless independent variable length {\tt L} 
is given by\\
{\tt L/(nStepsMaxCon + L*nStepsMaxVar)}.
(Defaults: 10000, 10000)
\end{itemize}
The step size is not allowed
to increase by more 
than a factor of 1.5 or 
decrease by more than a factor of 2 after each step or attempted step. 
Note that by 
setting {\tt precisionGoal} to 0,
one can arrange that the total number of steps on each leg tends to
{\tt nStepsMaxCon + L*nStepsMaxVar}. If instead one sets
{\tt precisionGoal} to a very large number, the number of steps will
tend to {\tt nStepsMin}.
The default values have been found to give good results for a wide variety
of different choices of input parameters, for the integration variables
used in the program.

In addition to the evaluation for generic parameters 
described above, {\tt TSIL}
provides functions for direct analytical evaluation of the vacuum
integrals $A(x)$ and $I(x,y,z)$, the one-loop integral $B(x,y)$, as
well as  $A(x^\prime),$ $B(x^\prime,y),$ $\partial B(x,y)/\partial s$, 
$A_\epsilon(x),$ $B_\epsilon(x,y),$
$I(x^\prime,y,z)$, $I(x'',y,z)$, $I(x',y',z)$, $I(x''',y,z)$,
and all $S$, $T$, $\overline T$, $U$, $V$, $M$ functions for which
results in terms of polylogarithms are available in the literature.

\section{How to use the program}
\label{sec:howto}
\setcounter{equation}{0}
\setcounter{footnote}{1}

\subsection{Building {\tt TSIL}}
\label{subsec:buildingTSIL}

Complete instructions for building the library are provided with the
distribution.  Typically the user edits the {\tt Makefile} to choose
the desired data size and set appropriate optimization flags for the
compiler. The command {\tt make} 
then produces the static archive {\tt libtsil.a},
which may be placed in any convenient directory {\tt <dir>}.  
The user then typically links to this library by passing the flags
$$
\mbox{{\tt -L<dir> -ltsil -lm -DTSIL\_SIZE\_<size>}}
$$
to the linker,
where the same flag {\tt -DTSIL\_SIZE\_<size>} was used in the {\tt Makefile} 
when compiling {\tt libtsil.a}.
The header file {\tt tsil.h} must be included in any
source file that makes use of the {\tt TSIL} routines or data
structures.

The command {\tt make} also produces an executable ${\tt tsil}$,
which takes as command-line arguments $x,y,z,u,v,s,Q^2$ and
prints out the values of all integral functions together with timing
and other information.
Also included with the package
is a test program {\tt testprog.c} (with
executable {\tt tsil.tst} produced by {\tt make}) and a 
set of 320 files containing comprehensive test data.  These include cases
representing all known analytic results as well as cases requiring
integration that have thresholds and pseudo-thresholds at $s=0$ and at
the final $s$.  Users should run this test suite after building the
library to insure that accurate results are obtained.  The test
program uses pass/fail/warn criteria that are controlled by macros
{\tt TSIL\_PASS} and {\tt TSIL\_WARN} in {\tt tsil\_testparams.h}.  
The first
sets the maximum relative error allowed for the test to pass; the
second sets a lower error threshold below which the test is deemed to
fail.  A relative error between these two values results in a warning.

\subsection{Essential functionality}

In the simplest application of {\tt TSIL}, 
the parameters 
$x,y,z,u,v \geq 0$ and $Q^2 > 0$ 
will be set using {\tt TSIL\_SetParameters}, 
the integrals for real $s$
evaluated using {\tt TSIL\_Evaluate}, and the results 
extracted
by the calling program with the command {\tt TSIL\_GetFunction}.
The code for this might look like:
\begin{flushleft}
~~~{\tt TSIL\_SetParameters (\&foo, x, y, z, u, v, qq);}\\
~~~{\tt TSIL\_Evaluate (\&foo, s);}\\
~~~{\tt integral1 = TSIL\_GetFunction (\&foo, <string1>);}\\
~~~{\tt integral2 = TSIL\_GetFunction (\&foo, <string2>);}\\
~~~{\tt \ldots}
\end{flushleft}
where {\tt foo} has type {\tt TSIL\_DATA}, and
{\tt x,y,z,u,v,s,qq} all have type {\tt TSIL\_REAL}, and
{\tt integral1}, {\tt integral2}, \ldots 
have type {\tt TSIL\_COMPLEX}, and {\tt <string1>}, {\tt <string2>}, \ldots 
can each be one of
\beq
&&
\mbox{\tt "M",~~~"Uzxyv",~~~"Uuyxv",~~~"Uxzuv",~~~"Uyuzv",~~~"Tvyz",} 
\nonumber \\
&&
\mbox{\tt "Tuxv",~~~"Tyzv",~~~"Txuv",~~~"Tzyv",~~~"Tvxu",~~~"Svyz",~~~"Suxv",}
\phantom{xxxxx}
\nonumber 
\eeq
according to which of the integrals in eq.~(\ref{eq:basis}) is 
desired. 
In addition, permutations
of the above, matching the symmetries of the basis
functions, are permitted.  Thus, for example, one can access
$U(z,x,y,v)$ by specifying either {\tt "Uzxyv"} or {\tt "Uzxvy"} in a call to
{\tt TSIL\_GetFunction}, since the $U$ functions are symmetric under
interchange of their last two arguments.  Likewise, any of {\tt "Suxv"},
{\tt "Sxuv"}, {\tt "Suvx"}, {\tt "Svux"}, {\tt "Sxvu"}, or {\tt "Svxu"} 
will return the function
$S(u,x,v)$ (symmetric under interchange of any of its arguments), etc.

Identifier strings can also be one of
\begin{center}
\mbox{\tt "Vzxyv",~~~"Vuyxv",~~~"Vxzuv",~~~"Vyuzv",~~~"Bxz",~~~"Byu"}, 
\end{center}
(and allowed permutations thereof) 
to access the functions $V$ and the one-loop $B$ functions.
Examples are given in subsection \ref{subsec:samples}. 
The functions ${\bf S}$, ${\bf T}$, ${\bf U}$ and ${\bf V}$ can
be accessed in a similar way, e.g.:
\begin{flushleft}
~~~{\tt integral3 = TSIL\_GetBoldFunction (\&foo, <string3>, n);}\\
\end{flushleft}
would return the coefficient of $1/\epsilon^n$ (for $n=0,1,$ or 2) 
in the bold-faced function corresponding to an appropriate
{\tt <string3>} from the list above.

All integrals that are analytically known in terms of polylogarithms
can also be evaluated directly, without {\tt TSIL\_SetParameters}
or {\tt TSIL\_Evaluate}
or {\tt TSIL\_GetFunction}. For example,
\begin{flushleft}
~~~{\tt TSIL\_Manalytic (x,y,z,u,v,s,\&result);}
\end{flushleft}
will return the {\tt int} value 1 and set the variable 
{\tt result} equal to $M(x,y,z,u,v)$ for the appropriate $s$ if it 
is analytically available, and otherwise
will return 0. Here {\tt x,y,z,u,v} are of type {\tt TSIL\_REAL},
and {\tt s, result} must be of type {\tt TSIL\_COMPLEX}. The functions
{\tt TSIL\_Sanalytic}, {\tt TSIL\_Tanalytic}, {\tt TSIL\_Tbaranalytic}, 
{\tt TSIL\_Uanalytic}, and {\tt TSIL\_Vanalytic} have analogous behavior,
except that they carry an additional argument {\tt qq} of type
{\tt TSIL\_REAL} for the renormalization scale squared $Q^2$. For example,
\begin{flushleft}
~~~{\tt TSIL\_Uanalytic (x,y,z,u,s,qq,\&result)}
\end{flushleft}
will return the {\tt int} value 1 and set the variable {\tt result} equal 
to $U(x,y,z,u)$ for the appropriate $s$ and $Q^2$, if it is 
analytically available, and otherwise will return 0.
The other analytic functions assign without pointers, for example
\begin{flushleft}
~~~{\tt result = TSIL\_Bp (x,y,s,qq);}
\end{flushleft}
will set {\tt result} equal to $B(x',y)$ computed analytically for
the appropriate $s$ and $Q^2$. Some examples are given in
the next subsection, and a complete list of the 
TSIL user API is given in 
Appendix \ref{appendixC} and the program documentation and header files.
  
In some applications, it could be that rather than
a master integral $M$ and all of its
subordinates, one may only need integral functions corresponding to
$S,T,U$ or only $S,T$. Those cases can be evaluated simply by
choosing any convenient numbers for the irrelevant squared-mass arguments.
However, this is clearly not optimally efficient. In version 1.1 of
TSIL, we have added the capability to only compute the integrals
in the $S,T,U$ system, or only those in the $S,T$ system.
This has been accomplished by adding functions 
{\tt TSIL\_SetParametersSTU} and {\tt TSIL\_SetParametersST}, 
each of which sets 
appropriate subsets of the squared mass parameters. A subsequent call of the function {\tt TSIL\_Evaluate} will only evaluate the relevant subset of integral
functions. So, the user code might include for example:
\begin{flushleft}
~~~{\tt TSIL\_SetParametersSTU (\&foo, x, z, u, v, qq);}\\
~~~{\tt TSIL\_Evaluate (\&foo, s);}\\
~~~{\tt integral1 = TSIL\_GetFunction (\&foo, <string1>);}\\
~~~{\tt integral2 = TSIL\_GetFunction (\&foo, <string2>);}\\
~~~{\tt \ldots}
\end{flushleft}
where {\tt <string1>}, {\tt <string2>}, \ldots 
can each be one of
\beq
&&
\mbox{\tt "Uxzuv",~~~"Tuxv",~~~"Txuv",~~~"Tvxu",~~~"Suxv",~~~"Bxz".}
\phantom{xxxxx}
\nonumber 
\eeq
Or, if $U(x,z,u,v)$ is not needed, then one can use  
{\tt TSIL\_SetParametersST (\&foo, x, u, v, qq);} instead.
For generic cases, the $S,T,U$ evaluation is a factor of $4$ or $5$ faster than
full evaluation, and the $S,T$ case gains a further 20\% in evaluation speed.
Note that the choice of which integral functions are evaluated by 
{\tt TSIL\_Evaluate} is controlled by the most recent call of
{\tt TSIL\_SetParameters} or {\tt TSIL\_SetParametersSTU} or
{\tt TSIL\_SetParametersST} for the relevant data struct.
Also, if $S,T,U$ or $S,T$ evaluation is used, then 
{\tt TSIL\_GetData} and {\tt TSIL\_GetBoldData}
will generate an error message; only
{\tt TSIL\_GetFunction} and {\tt TSIL\_GetBoldFunction}
should be used to extract the results in those cases.
Note that in subset evaluation cases where there is only a single function 
of a given type ($U$, $V$, $S$, or $B$), the specification string may be 
abbreviated to only the first
character, e.g. ``{\tt U}'' in place of ``{\tt Uxzuv}'' above.

In version 1.2 of TSIL, we have introduced a new struct type 
{\tt TSIL\_RESULT}, which contains only the subset of the information 
found in {\tt TSIL\_DATA} that is essential in typical applications. 
This information consists of: the squared masses $x,y,z,u,v$,  
the external momentum invariant $s$, the renormalization scale $Q^2$
and the integral results listed in eq.~(\ref{eq:basis}) as well as
$\overline T(v,y,z)$, $\overline T(u,x,v)$, $\overline T(y,z,v)$,
$\overline T(z,y,v)$, $B(x,z)$ and $B(y,u)$. The {\tt TSIL\_RESULT} 
struct is useful for more efficient storage of results and for performing 
permutations of the squared masses. The new function 
{\tt TSIL\_CopyResult} can be used to copy data from a full 
{\tt TSIL\_DATA} struct to a {\tt TSIL\_RESULT} struct. 
(This should be done only after {\tt TSIL\_Evaluate} has been called on the
{\tt TSIL\_DATA} struct.) The new function 
{\tt TSIL\_PermuteResult} can then be used to copy the data from one 
{\tt TSIL\_RESULT} struct to another, with the option of permuting the 
squared masses according to either $(x,z)\leftrightarrow (y,u)$, or 
$(x,y)\leftrightarrow (z,u)$, or $(x,y)\leftrightarrow (u,z)$, by using 
the symmetries of the master topology rather than needlessly redoing 
the integrals. Such permutations occur often in practical 
applications such as the supersymmetric Standard Model. 

In addition, functions {\tt TSIL\_GetFunctionR} and {\tt TSIL\_GetDataR} (new in v1.4) have
the same functionality as {\tt TSIL\_GetFunction} and {\tt TSIL\_GetData}, but with a {\tt TSIL\_RESULT}
argument instead of {\tt TSIL\_DATA.}

The elements of
the {\tt TSIL\_DATA} and {\tt TSIL\_RESULT} struct types can be found in 
the file {\tt tsil.h}.

\subsection{Sample applications}
\label{subsec:samples}

As a sample application of {\tt TSIL}, let us calculate the
two-loop self energy and pole squared mass for a single scalar field
with interaction Lagrangian
\beq
{\cal L} = -{1\over 2}m^2\phi^2 - {g\over 3!}\phi^3 
- {\lambda\over 4!}\phi^4\; .
\label{eq:intlag}
\eeq
Here $m^2$, $g$ and $\lambda$ are the tree-level renormalized
parameters.  The self-energy 
\beq
\Pi(s)={1\over 16\pi^2}\Pi^{(1)}(s) + {1\over (16\pi^2)^2}\Pi^{(2)}(s)
+\cdots
\eeq
is a function of $s=-p^2$, with $p^\mu$ the external momentum, as well
as the parameters $m^2$, $g$ and $\lambda$.  Note that the metric is
either of signature ($-$$+$$+$$+$) or Euclidean.  Furthermore, $s$ must be
taken to be real with an infinitesimal positive imaginary part to
properly resolve the branch cuts.

The pole squared mass 
\beq
s=M^2-i\Gamma M
\eeq
is defined as the position of the pole, with non-positive imaginary
part, in the propagator obtained from the perturbative Taylor
expansion of the self-energy about the tree-level squared
mass.\footnote{In a theory with gauge fields the self-energy is
gauge-dependent, but the pole squared mass defined in this way is
properly gauge invariant.}  (Note that in the present case the width 
$\Gamma=0$
identically.)  This leads to the following iterative scheme for
computing the two-loop pole squared mass.  First, the one-loop
approximation $s^{(1)}$ to the pole squared mass is obtained as
\beq
s^{(1)} = m^2 + {1\over16\pi^2} \Pi^{(1)}(m^2)\; .
\label{eq:step1}
\eeq
Then, defining
\beq
\tilde{\Pi} = 
{1\over 16\pi^2}\left[ \Pi^{(1)}(m^2) + (s^{(1)}-m^2)
{\Pi^{(1)\prime}}(m^2)\right]
 + {1\over (16\pi^2)^2}\Pi^{(2)}(m^2)\; ,
\eeq
in which the prime indicates a derivative with respect to $s$, we obtain
the two-loop approximation to the pole squared mass as
\beq
s^{(2)} = m^2 + \tilde{\Pi}\; .
\label{eq:step2}
\eeq 

For the scalar theory described by eq. (\ref{eq:intlag}) we find,
including the $\overline{\rm MS}$ counterterms: 
\bea
\Pi^{(1)}(s) &=& {1\over2}\lambda A(x) - {1\over2}g^2 B(x,x)\\
\Pi^{(2)}(s) &=& -{1\over2}g^4 M(x,x,x,x,x) 
-{1\over2}g^4V(x,x,x,x) +\lambda g^2U(x,x,x,x)  
\nonumber\\
&& -{1\over6}\lambda^2 S(x,x,x)
+{1\over4}\lambda g^2 B(x,x)B(x,x)
+{1\over4}\lambda^2 A(x)\left[A(x)/x+1\right]\\
&& - {1\over2}\lambda g^2 A(x)B(x^\prime,x)
- {1\over4}\lambda g^2 I(x^\prime,x,x)\; ,
\nonumber
\eea
where $x=m^2$, and 
\beq
\Pi^{(1)\prime} (s) 
= -{1\over 2} g^2 \frac{\partial}{\partial s} B(x,x)
\; .
\eeq
Note that both $\Pi^{(1)\prime}$ and $\Pi^{(2)}$ have $({1-s/4x})^{-1/2}$
singularities at the threshold $s=4x$, though this is not manifest in
the above formulas.  The {\tt TSIL} code handles such true
singularities by returning a value that is interpreted and displayed
as the string {\tt ComplexInfinity}.

Below is a sample C code that uses {\tt TSIL} to calculate the
pole squared mass for parameter values $m^2$, $g$, $\lambda$ and $Q^2$
obtained as command-line inputs in that order.  It first
computes the required basis functions at $s=m^2$, then assembles
$\Pi^{(1)}(m^2)$, $\Pi^{(2)}(m^2)$ and $\Pi^{(1)\prime}(m^2)$, 
and finally outputs the pole squared mass:

{\small
\begin{verbatim}
/*   === scalarpole.c ===
 *
 *   Command-line arguments are: 
 *          scalar mass squared = x, 
 *          cubic coupling = g, 
 *          quartic coupling = lambda, 
 *          renormalization scale squared = qq. 
 *
 *    Run as, for example: ./spole 1 2 3 1
 */

#include <stdio.h>
#include "tsil.h"  /* Required TSIL header file */

#define PI 4.0L*TSIL_ATAN(1.0L) /* Uses arctan function defined in tsil.h */

int main (int argc, char *argv[])
{
  TSIL_DATA    result; /* Top-level TSIL data object */
  TSIL_REAL    qq;     /* Ensures correct basic type; see also TSIL_COMPLEX */ 
  TSIL_REAL    x, g, lambda;
  TSIL_COMPLEX pi1, pi1prime, pi2, s1, s2;
  TSIL_REAL    factor = 1.0L/(16.0L*PI*PI);

  /* If incorrect number of args, print message on stderr and exit: */
  if (argc != 5)
    TSIL_Error("main", "Expected 4 arguments: m^2, g, lambda, and Q^2", 1);

  /* Note cast to appropriate floating-point type for safety */
  x      = (TSIL_REAL) strtold(argv[1], (char **) NULL); 
  g      = (TSIL_REAL) strtold(argv[2], (char **) NULL);
  lambda = (TSIL_REAL) strtold(argv[3], (char **) NULL); 
  qq     = (TSIL_REAL) strtold(argv[4], (char **) NULL); 

  /* All loop integrals have a common squared-mass argument x: */
  TSIL_SetParameters (&result, x, x, x, x, x, qq); 

  /* For the pole mass calculation, evaluate two-loop integrals at s = x: */
  TSIL_Evaluate (&result, x);

  /* Assemble one- and two-loop mass squared results: */
  pi1 = 0.5L*lambda*TSIL_A(x,qq) - 0.5L*g*g*TSIL_B(x,x,x,qq);

  pi1prime = -0.5L*g*g*TSIL_dBds(x, x, x, qq); 

  pi2 = - 0.5L*g*g*g*g*TSIL_GetFunction(&result, "M")
        - 0.5L*g*g*g*g*TSIL_GetFunction(&result, "Vzxyv")
               + lambda*g*g*TSIL_GetFunction(&result, "Uzxyv")
    - (1.0L/6.0L)*lambda*lambda*TSIL_GetFunction(&result, "Svyz")
    + 0.25L*lambda*g*g*TSIL_POW(TSIL_GetFunction(&result, "Bxz"), 2)
    + 0.25L*lambda*lambda*TSIL_A(x,qq)*(TSIL_A(x,qq)/x + 1.0L)   
    - 0.5L*lambda*g*g*TSIL_A(x,qq)*TSIL_Bp(x, x, x, qq)      
    - 0.25L*lambda*g*g*TSIL_I2p(x,x,x,qq);               

  s1 = x + factor*pi1;
  s2 = x + factor*pi1 + factor*factor*(pi2 + pi1*pi1prime);

  printf("Tree-level squared mass:    %lf\n", (double) x); 
  printf("One-loop pole squared mass: %lf\n", (double) s1); 
  printf("Two-loop pole squared mass: %lf\n", (double) s2); 

  return 0;
}
\end{verbatim}
}
\noindent
Note the use of {\tt TSIL\_A}, {\tt TSIL\_I2p}, {\tt TSIL\_B},
{\tt TSIL\_Bp}, and {\tt TSIL\_dBds} to evaluate the functions
$A(x)$, $I(x',x,x)$, $B(x,x)$, $B(x',x)$, and $\partial B(x,x)/\partial s$,
respectively. [In the evaluation of {\tt pi2}, we arbitrarily chose to use
{\tt TSIL\_GetFunction(\&result, "Bxz")} where 
{\tt TSIL\_B} could have been used.]
The compile command for this program to produce the executable
{\tt spole} would typically be
\begin{center}
{\tt cc -o spole scalarpole.c -L<dir> -ltsil -lm -DTSIL\_SIZE\_<size>}
\end{center}
as discussed in subsection \ref{subsec:buildingTSIL}.

\begin{figure}[t]
\begin{center}
\begin{picture}(100,75)(-50,-27)
\SetWidth{0.9}
\DashLine(-50,0)(-22,0){4}
\DashLine(50,0)(22,0){4}
\CArc(0,0)(22,-90,90)
\DashCArc(0,0)(22,90,270){4}
\Line(0,-22)(0,22)
\Text(5,0)[]{$\tilde g$}
\Text(-21,23.8)[]{$\tilde t_m$}
\Text(-20.5,-19)[]{$\tilde t_n$}
\Text(18.5,21)[]{$t$}
\Text(18.5,-19.5)[]{$t$}
\Text(-47,7)[]{$h^0$}
\end{picture}
\hspace{1.0cm}
\begin{picture}(100,75)(-50,-27)
\SetWidth{0.9}
\DashLine(-50,-18)(-22,-18){4}
\DashLine(50,-18)(22,-18){4}
\Line(-22,-18)(22,-18)
\DashCArc(0,18)(22,0,180){4}
\Line(-22,18)(22,18)
\Line(-22,-18)(-22,18)
\Line(22,-18)(22,18)
\Text(0,-24.5)[]{$t$}
\Text(-27,0)[]{$t$}
\Text(27,0)[]{$t$}
\Text(0,12)[]{$\tilde g$}
\Text(1,48.5)[]{$\tilde t_m$}
\Text(-47,-11.5)[]{$h^0$}
\end{picture}
\hspace{1.0cm}
\begin{picture}(100,75)(-50,-27)
\SetWidth{0.9}
\DashLine(-50,-18)(-22,-18){4}
\DashLine(50,-18)(22,-18){4}
\DashLine(-22,-18)(22,-18){4}
\CArc(0,18)(22,0,180)
\Line(-22,18)(22,18)  
\DashLine(-22,-18)(-22,18){4}
\DashLine(22,-18)(22,18){4}
\Text(0,-26.5)[]{$\tilde t_m$}
\Text(-28,0)[]{$\tilde t_n$}
\Text(28.5,0)[]{$\tilde t_k$}
\Text(0,12)[]{$t$}
\Text(1,48.5)[]{$\tilde g$}
\Text(-46,-11.5)[]{$h^0$}
\end{picture}
\end{center}
\caption{\label{fig:higgs} 
Feynman diagram topologies for some two-loop corrections to the neutral Higgs
scalar boson self-energies in supersymmetry.}
\end{figure}
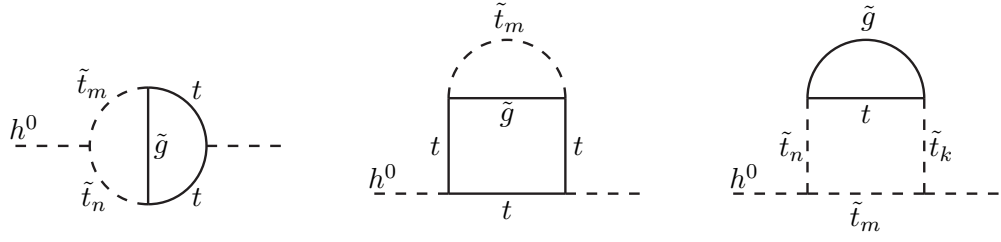

Other situations can be treated with appropriate generalizations. For
example, in the Minimal Supersymmetric Standard Model correction 
to the neutral Higgs scalar boson 
self-energies \cite{Martin:2004kr}, 
there are graphs with the topology shown in fig.~\ref{fig:higgs}
involving the top quark $t$, the top squarks $\tilde t_n$ for $n=1,2$,
and the gluino $\tilde g$. 
All of the necessary one-loop and two-loop basis integrals for
these diagram topologies can be computed in one fell swoop with:
\begin{flushleft}
{\tt TSIL\_DATA result[2][2];  /* Declare an array of TSIL\_DATA structs */}\\
{\tt TSIL\_REAL mt2, mstop2[2], mgluino2;
/* Squared masses of top, stops, gluino */}\\
{\tt TSIL\_REAL s;}\\ 
{\tt int i,j;}\\
{\tt \ldots}\\
{\tt for (i=0; i<2; i++) \{}\\
~~~{\tt for (j=0; j<i; j++) \{}\\
~~~~~{\tt TSIL\_SetParameters~(\&(result[i][j]),mstop2[i],mt2,%
mstop2[j],mt2,mgluino2,qq);}\\
~~~~~{\tt TSIL\_Evaluate (\&(result[i][j]),s);}\\
~~~{\tt \}}\\
{\tt \}}
\end{flushleft}
The index of {\tt mstop2[]} in the code 
is one less than the conventional
top squark mass eigenstate label, because arrays start at index 0 in C.
Note that the evaluation for $j=1$, $i=0$ would be redundant by symmetry
with that for $j=0$, $i=1$, and so is not performed here.
Note also that the subordinate integrals of topology $S$, $T$, and $U$ are 
needed for the calculation of the self-energy and the pole mass, 
even though there are no Feynman diagrams
with the corresponding topologies since there are no
four-particle couplings involving fermions. The necessary two-loop
integrals can then be extracted by, for example:
\begin{flushleft}
~~~{\tt value = TSIL\_GetFunction(\&(result[0][0]), "M");}
\hfill 
for $M(m_{\tilde t_1}^2,m_t^2,m_{\tilde t_1}^2, m_t^2, m_{\tilde g})$,
\\
~~~{\tt value = TSIL\_GetFunction(\&(result[0][0]), "Vuyxv");} 
\hfill 
for $V(m_t^2,m_t^2, m_{\tilde t_1}^2, m_{\tilde g})$,
\\
~~~{\tt value = TSIL\_GetFunction(\&(result[1][0]), "Vzxyv");} 
\hfill
for $V(m_{\tilde t_1}^2, m_{\tilde t_2}^2, m_t^2, m_{\tilde g})$,
\\
~~~{\tt value = TSIL\_GetFunction(\&(result[1][0]), "Vxzuv");} 
\hfill
for $V(m_{\tilde t_2}^2, m_{\tilde t_1}^2, m_t^2, m_{\tilde g})$,
\\
~~~{\tt value = TSIL\_GetFunction(\&(result[0][0]), "Txuv");} 
\hfill
for $T(m_{\tilde t_1}^2, m_t^2, m_{\tilde g})$,
\end{flushleft}
etc. 

For other examples of the usage of {\tt TSIL}, see the 
public code {\tt SMH} \cite{Martin:2014cxa}, which calculates the 
Standard Model Higgs pole mass including all 2-loop order corrections,
and the implementation of 2-loop gluino and squark masses in {\tt SOFTSUSY3.7}
\cite{Allanach:2016rxd}.

\subsection{Using {\tt TSIL} with C++}
\label{sec:cplusplus}

{\tt TSIL} functions can be called from C++ code using {\tt libtsil.a}.  
The header file {\tt tsil\_cpp.h} 
should be included in any C++ source files that make use of {\tt TSIL}
functions.  This file is equivalent to the usual {\tt tsil.h}, but with
additional definitions to streamline interoperation with C++.

In particular, {\tt tsil\_cpp.h} provides wrappers for {\tt TSIL} 
functions that insure proper handling of complex values, which are of 
generic type {\tt \_Complex} in C and {\tt std::complex<>} in C++.  The 
relevant standards guarantee that pointers to these types will be 
interpreted correctly in any context, and the wrappers insure that 
complex arguments and return values are always passed between C and C++ 
in this way.

What this means for the user is that the {\tt TSIL} functions all have
C++-specific versions that can be called with C++ types as arguments
and will return C++ types.  The names of these are the same as the
corresponding TSIL functions, with a trailing underscore.  Thus the C
function
\begin{flushleft}
{\tt TSIL\_GetFunction (...)}
\end{flushleft}
becomes
\begin{flushleft}
{\tt TSIL\_GetFunction\_ (...) }
\end{flushleft}
when called from C++, etc.  All functions in the user API have been
supplied with such wrappers, even though not all functions really need
them; this is so that users need not remember which functions have
special names.

See the {\tt TSIL} documentation file {\tt README.txt}
for additional details on using {\tt TSIL} with C++.

\subsection{Using {\tt TSIL} with FORTRAN}
\label{sec:fortran}

{\tt TSIL} functions can be called from Fortran
programs, and utilities for this are included with the package.  
Basic functionality is provided by a ``wrapper'' function
{\tt tsilfevaluate}, which is called as a subroutine from 
a Fortran program.
This subroutine implements the most general {\tt TSIL} calculation: 
it takes as arguments $x,y,z,u,v,Q^2,s$ and returns the values of all 
basis functions, including $\overline{T}$, $V$, and ``bold'' functions.

The results are returned to the calling program in a {\tt COMMON}
block, which corresponds to a special C {\tt struct} used in {\tt
tsilfevaluate}.  This {\tt COMMON} block contains a number of pre-defined
arrays that hold the various function values.  
Definitions of the {\tt COMMON} block and subsidiary arrays are given in
a header file that users include in their Fortran programs 
({\tt tsil\_fort.inc}).  In addition,
this header file defines a set of integer variables that
allow items in the {\tt COMMON} block to be referred to by name.

A Fortran program fragment that uses these utilities is shown below:

{\small
\begin{verbatim}
       PROGRAM ftest
c      Includes array and COMMON definitions:
       INCLUDE 'tsil_fort.inc'

c      (Code setting values of x,y,z,u,v,qq,s not shown)
       ...
c      Evaluate basis integrals:
       CALL tsilfevaluate(x,y,z,u,v,qq,s)

c      Print a representative value:
       PRINT *, U(xzuv)
       ...
\end{verbatim}
}

The provided wrapper code can serve as a model for users wishing to
write their own interface routines with additional functionality.
The TSIL documentation contains a detailed discussion of the issues
that arise in using TSIL with Fortran.

\section{Outlook}
\label{sec:outlook}
\setcounter{equation}{0}
\setcounter{footnote}{1}

The {\tt TSIL} library is intended to provide a convenient all-purpose
solution to the problem of numerical evaluation of two-loop self-energy
integrals in high-energy physics. The functions provided should work for
arbitrary input parameters, without relying on special mass hierarchies or
other simplifications that can arise in special cases. 
The library does take advantage of simplifications when they allow
evaluation in terms of polylogarithms. 
 
The library is organized around the calculation of the basis integrals to
which all other self-energy contributions can be reduced by known
algorithms.  It is therefore the responsibility of the user to perform the
non-trivial calculations necessary to assemble the basis functions into
physical observables. This involves the reduction of the Feynman diagrams
to the basis integrals and the inclusion of counterterms, tasks that can
always be automated using purely symbolic computer algebra manipulations
or even performed by hand in favorable situations. We believe that keeping
these separate from the problem of numerical evaluation 
is advantageous, and that the modular nature of
this approach will afford the flexibility to deal with the surprises that
hopefully await us as we explore physics at the TeV scale.

\addcontentsline{toc}{section}{Appendix A: Comparison of conventions
with other sources}
\section*{Appendix A: Comparison of conventions with other sources}
\label{appendixA} \renewcommand{\theequation}{A.\arabic{equation}}
\setcounter{equation}{0}
\setcounter{footnote}{1}

In this Appendix, we list the correspondence between our notation for the
loop integrals and those found in several other references. In the
following, the functions as defined in this paper and in
refs.~\cite{Martin:2003qz},\cite{Martin:2003it} always appear on the 
left-hand side, and
equivalent notations for them in other papers appear on the right. 

The definition of the master integral in ref.~\cite{Broadhurst:1987ei}
is given by:
\beq
M(x,y,z,u,v) 
&=& -I(s)/s,
\eeq
with $m_1^2, m_2^2, m_3^2, m_4^2, m_5^2 = x,z,v,y,u$, and the integral
on the right-hand side defined in 4 dimensions. 

The notation used in
refs.~\cite{%
     Weiglein:hd,
     Scharf:1993ds,
     Berends:1994ed,
     Bauberger:1994nk,
     Bauberger:1994hx} is:
\beq
{\bf A}(x) &=& -A_0(x)
\\
{\bf B}(x,y) &=& B_0(s;x,y)
\\
{\bf I}(x,y,z) &=& -T_{135}(x,y,z)
\\
{\bf S}(x,y,z) &=& -T_{234}(s;x,y,z)
\\
{\bf T}(x,y,z) &=& T_{2234}(s;x,x,y,z)
\\
{\bf U}(x,y,z,u) &=& T_{1234}(s;y,x,z,u)
\\
{\bf V}(x,y,z,u) &=& -T_{11234}(s;y,y,x,z,u)
\\
{\bf M}(x,y,z,u,v) &=& -T_{12345}(s;x,z,v,u,y) .
\eeq
Ref.~\cite{Bauberger:1994by} used the same notation, with the
exception that ${\bf S}(x,y,z) = -T_{123}(s;x,y,z)$ there.

The notation in ref.~\cite{Tarasov:1997kx} is:  
\beq
{\bf A}(x) &=& i \cTarasov T^{(d)}_{1}
\quad\>\, {\rm with}\>\,
m_1^2
= x,
\\
{\bf B}(x,y) &=& -i\cTarasov G^{(d)}_{11}(s)
\quad\>\, {\rm with}\>\,
m_1^2,
m_2^2
= x,y ,
\\
{\bf I}(x,y,z) &=& \cTarasov^2 J^{(d)}_{111}(0)
\quad\>\, {\rm with}\>\,
m_1^2,
m_2^2,
m_3^2
= x,y,z ,
\\
{\bf S}(x,y,z) &=& \cTarasov^2 J^{(d)}_{111}(s)
\quad\>\, {\rm with}\>\,
m_1^2,
m_2^2,
m_3^2
= x,y,z ,
\\
{\bf T}(x,y,z) &=& -\cTarasov^2 J^{(d)}_{211}(s)
\quad\>\, {\rm with}\>\,
m_1^2,
m_2^2,
m_3^2
= x,y,z ,
\\
{\bf U}(x,y,z,u) &=& -\cTarasov^2 V^{(d)}_{1111}(s) 
\quad\>\, {\rm with}\>\,
m_1^2,
m_2^2,
m_3^2,
m_4^2
= u,x,z,y ,
\\
{\bf V}(x,y,z,u) &=& \cTarasov^2 V^{(d)}_{1112}(s)
\quad\>\, {\rm with}\>\,
m_1^2,
m_2^2,
m_3^2,
m_4^2
= u,x,z,y ,
\\
{\bf M}(x,y,z,u,v) &=& \cTarasov^2 F^{(d)}_{11111}(s) 
\quad\>\, {\rm with}\>\,
m_1^2,
m_2^2,
m_3^2,
m_4^2,
m_5^2 = x,y,z,u,v ,
\eeq
and the closely related notation of ref.~\cite{Mertig:1998vk} is
\beq
{\bf A}(x) &=& i \cTarasov \mbox{{\tt TAI$[$d,s,$\{\{1,\sqrt{x}\}\}]$}},
\\
{\bf B}(x,y) &=& -i \cTarasov \mbox{{\tt TBI$[$d,s,$\{%
\{1,\sqrt{x}\},\{1,\sqrt{y}\}\}]$}} 
\\
{\bf I}(x,y,z) &=& \cTarasov^2 \mbox{{\tt TJI$[$d,0,$\{%
\{1,\sqrt{x}\},%
\{1,\sqrt{y}\},%
\{1,\sqrt{z}\}%
\}]$}}
\\
{\bf S}(x,y,z) &=& \cTarasov^2 \mbox{{\tt TJI$[$d,s,$\{%
\{1,\sqrt{x}\},%
\{1,\sqrt{y}\},%
\{1,\sqrt{z}\}%
\}]$}}
\\
{\bf T}(x,y,z) &=& -\cTarasov^2 \mbox{{\tt TJI$[$d,s,$\{%
\{2,\sqrt{x}\},%
\{1,\sqrt{y}\},%
\{1,\sqrt{z}\}%
\}]$}}
\\
{\bf U}(x,y,z,u) &=& -\cTarasov^2 \mbox{{\tt TVI$[$d,s,$\{%
\{1,\sqrt{u}\},%
\{1,\sqrt{x}\},%
\{1,\sqrt{z}\},%
\{1,\sqrt{y}\}%
\}]$}}
\\
{\bf V}(x,y,z,u) &=& \cTarasov^2 \mbox{{\tt TVI$[$d,s,$\{%
\{1,\sqrt{u}\},%
\{1,\sqrt{x}\},%
\{1,\sqrt{z}\},%
\{2,\sqrt{y}\}%
\}]$}}
\\
{\bf M}(x,y,z,u,v) &=& \cTarasov^2 \mbox{{\tt TFI$[$d,s,$\{%
\{1,\sqrt{x}\},%
\{1,\sqrt{y}\},%
\{1,\sqrt{z}\},%
\{1,\sqrt{u}\},%
\{1,\sqrt{v}\}%
\}]$}} 
\phantom{xxxx}
\eeq
where 
$
\cTarasov = (4 \pi \mu^2)^{2-d/2}.
$

The notation of 
\cite{%
Caffo:1999nk,
Caffo:1998du,
Caffo:2002ch,
Caffo:2002wm,
Caffo:2003ma}
(ref.~\cite{Caffo:1998yd} used a slightly different normalization)
is:
\beq
{\bf A}(x) &=& \cCCLR T(d,x) 
\\
{\bf B}(x,y) &=& \cCCLR S(d,x,y,-s) 
\\
{\bf I}(x,y,z) &=& \cCCLR^2 V(d,x,y,z) 
\\
{\bf S}(x,y,z) &=& \cCCLR^2 F_0(d,x,y,z,-s) 
\\
{\bf T}(x,y,z) &=& 
\cCCLR^2 F_1(d,x,y,z,-s) = 
\cCCLR^2 F_2(d,z,x,y,-s) = 
\cCCLR^2 F_3(d,y,z,x,-s) 
\\
{\bf U}(x,y,z,u) &=& \cCCLR^2 F_4(d,x,u,y,z,-s) ,
\eeq
where
$
\cCCLR = 4 \mu^{4-d} .
$

The notation of \cite{Passarino:2001jd} is:
\beq
{\bf S}(x,y,z) &=& \cPassarino^2 S^{111}
\qquad{\rm with}\>
m_1^2,
m_2^2,
m_3^2 = x,y,z,
\\
{\bf U}(x,y,z,u) &=& \cPassarino^2 S^{121}
\qquad{\rm with}\>
m_1^2,
m_2^2,
m_3^2,
m_4^2 = z,u,y,x,
\\
{\bf M}(x,y,z,u,v) &=& \cPassarino^2 S^{221}
\qquad{\rm with}\>
m_1^2,
m_2^2,
m_3^2,
m_4^2,
m_5^2 = x,z,v,y,u,
\phantom{xxx}
\eeq
with $\cPassarino = (2 \pi)^{d-4}$.

\addcontentsline{toc}{section}{Appendix B: Analytic expressions for some
special cases}
\section*{Appendix B: Analytic expressions for some
special cases}
\label{appendixB} \renewcommand{\theequation}{B.\arabic{equation}}
\setcounter{equation}{0}
\setcounter{footnote}{1}

In this Appendix, we present some analytic formulas for some important
two-loop basis integrals special cases, two of which do not seem to have 
appeared before in the literature, and others that are equivalent to known 
results.

As a generalization of the integral $M(0,0,x,x,0)$ computed in
ref.~\cite{Broadhurst:1987ei}, we find that for $x\geq y$:
\beq
M(0,0,x,y,0) &=& 
\Bigl [ 
3 \trilog (r_x) + 3 \trilog (r_y) 
-3 \trilog (r_y/r_x) - 3 \trilog (y r_y/x r_x) 
\nonumber \\ &&
+ 3 \trilog(y/x) - 3 \zeta(3) 
+ [\ln (r_y) - \ln (r_x) + 2 \ln(y/x)] \dilog (y r_y/x r_x)
\nonumber \\ &&
+ [\ln (r_y) - \ln (r_x) - 2 \ln(y/x)] \dilog (y/x)
+ [\ln (r_y) - 3 \ln (r_x)] \dilog (r_x)
\nonumber \\ &&
+ [\ln (r_x) - 3 \ln (r_y)] \dilog (r_y)
+ 3 [\ln (r_y) - \ln (r_x)] \dilog (r_y/r_x)
\nonumber \\ &&
+ [\ln(r_x) - \ln(r_y)][\ln (r_x) - \ln (r_y) + \ln (x/y)] \ln (1-y/x)
\nonumber \\ &&
- \ln^3 (r_x)
+ 2 \ln^2 (r_x) \ln (r_y)
- \ln (r_x) \ln^2 (r_y) 
- \ln (r_x) \ln (r_y) \ln (1-r_x)
\nonumber \\ &&
+ \ln (r_x) \ln (r_y) \ln (y/x)
- \ln (r_x) \ln (y/x) [\ln (r_x) + \ln (y/x)]/2
\nonumber \\ &&
- \zeta(2) [3 \ln (r_x) + \ln (r_y)]
\Bigr ]/s 
\eeq
where $r_x = 1-s/x$ and $r_y = 1-s/y$ are each given an infinitesimal
negative imaginary part. The result for $x<y$ is obtained 
by interchanging the squared masses.

Generalizing the result $M(0,x,x,0,x)|_{s=x}$ found in 
ref.~\cite{Broadhurst:1987ei}, we obtain threshold cases:
\beq
M(0,y,y,0,x)|_{s=x} &=& \Bigl [
2 \trilog ([r-1]/[r+1]) - 2 \trilog ([1-r]/[r+1])
-2 \trilog (r/[r+1]) 
\nonumber \\ &&
-2 \trilog (1-1/r) 
- \trilog (1/r^2)/4
+ 2 [\ln (2r) - \ln(r+1)] \dilog(r/[r+1])
\nonumber \\ &&
+ [4 \ln (r-1) - 2 \ln (2r) - 2 \ln (r+1)] \dilog (1-1/r)
\nonumber \\ &&
+ 2 [\ln(r-1) - \ln(r+1)][\dilog([1-r]/2) - \dilog ([r-1]/2r)
\nonumber \\ &&
- \ln (4r) \dilog(1/r^2)/2
+ \ln^2 (r) [3 \ln(r-1) - \ln(r+1) - 6 \ln 2]/2
\nonumber \\ &&
+ \ln (r) [2 \ln(r+1) \ln(2[r-1]) -\ln^2(r-1)- 6 \zeta (2) ]
-(4/3) \ln^3 (r)
\nonumber \\ &&
-(2/3) \ln^3 (r+1) 
+ \ln 2 \ln^2 (r+1)
+ \zeta (3)/2 + 6 \zeta(2) \ln(1+r)
\Bigr ]/x ,
\phantom{xxx}
\\
M(0,x,y,0,y)|_{s=x} &=& 
\Bigl [
2 \trilog ([\sqrt{r}-1]/[\sqrt{r} + 1])
-2 \trilog ([1-\sqrt{r}]/[\sqrt{r} + 1])
\nonumber \\ &&
+2 \trilog (1-r)
-2 \trilog ([r-1]/r)
-3 \zeta (3)/2
\nonumber \\ &&
+ [\ln (r) - 2 \ln(r-1)] \dilog(1-r)
- [\ln^2 (r-1) \ln (r)]/2
\nonumber \\ &&
+ [\ln^3 (r)]/3
-2 \zeta(2) \ln (r)
+ 6 \zeta (2) \ln (1 + \sqrt{r})
\Bigr ]/x,
\label{eq:Moxyoy}
\eeq
where $r = y/x$ is given an infinitesimal negative 
imaginary part to get the correct branches.
Eq.~(\ref{eq:Moxyoy}) is equivalent, by use of the recurrence relations 
of \cite{Tarasov:1997kx}, to results already obtained in 
\cite{Fleischer:1998dw,Jegerlehner:2003py}.

We also note the following threshold and pseudo-threshold values 
(equivalent to results obtained in \cite{Gray:1990yh}; see also
\cite{Davydychev:1998si}):
\beq
U(x,0,y,y)|_{s=x} &=& 11/2 - 3 \lnbar x
                    + \lnbar x \lnbar y - (\lnbar y)^2/2
                    + (1 + y/x) [ \zeta (2) - \dilog (1-x/y)]
\phantom{xx}
\nonumber \\ &&
                    -4 r \lbrace 
                    \dilog(
                    [ 1- r ]/ 
                    [ 1 + r ]) 
                    -\dilog(
                    [ r-1 ]/ 
                    [ r+1 ]) + 3 \zeta(2)/2 \rbrace 
\eeq
where $r = \sqrt{y/x}$, and
\beq
U(y,0,y,x)|_{s=x} &=& 
11/2 - 2\lnbar x - \lnbar y + (\lnbar y)^2/2
 -2 \zeta(2) (1 + y/x)
\nonumber \\ &&
 + (\lnbar x - 1 + y/x [1- \lnbar y]) \ln (1-x/y - i \varepsilon)
 + (1 + 2 y/x) \dilog(1-x/y)
.
\phantom{xxx}
\eeq

\addcontentsline{toc}{section}{Appendix C: The {\tt TSIL} Application 
Programmer Interface}
\section*{Appendix C: The {\tt TSIL} Application Programmer Interface}
\label{appendixC} \renewcommand{\theequation}{C.\arabic{equation}}
\setcounter{equation}{0}
\setcounter{footnote}{1}

This Appendix lists the functions in the {\tt TSIL}
package, and their basic functionality.  Complete details may be found
in the {\tt TSIL} documentation and header files.

\begin{tabbing}
{\tt TSIL\_GetBoldFunctionxxx}  
     \= Return a single specified bold basis function value\kill
\underline{Basic evaluation functions:} \> \\
{\tt TSIL\_SetParameters} 
     \> Sets parameters $x,y,z,u,v,Q^2$ 
     and selects evaluation of
     all \\ \>integral functions\\
{\tt TSIL\_SetParametersSTU} 
     \> Sets parameters $x,z,u,v,Q^2$ and selects evaluation of
     $S,T,U$ \\ \>functions only.~~(New in v1.1 November 2006; see section 5.2.)\\
{\tt TSIL\_SetParametersST} 
     \> Sets parameters $x,u,v,Q^2$ and selects evaluation of
     $S,T$ \\ \>functions only.~~(New in v1.1 November 2006; see section 5.2.)\\
{\tt TSIL\_Evaluate}
     \> Evaluate integral functions for a specified $s$\\
{\tt TSIL\_GetStatus} 
     \> Return current evaluation status\\
{\tt TSIL\_GetData} 
     \> Extract a set of integral function values to an array\\ 
{\tt TSIL\_GetBoldData} 
     \> Extract a set of bold integral function values to an array\\
{\tt TSIL\_GetFunction}
     \> Return a single specified integral function value\\
{\tt TSIL\_GetBoldFunction}  
     \> Return a single specified bold integral function value\\
{\tt TSIL\_GetDataR} 
     \> Extract a set of integral function values to an array, from a\\ 
     \> {\tt TSIL\_RESULT} struct. (New in v1.4 May 2016.)\\
{\tt TSIL\_GetFunctionR}
     \> Return a single specified integral function value from a \\
     \> {\tt TSIL\_RESULT} struct. (New in v1.4 May 2016.)\\
\end{tabbing}

\begin{tabbing}
\underline{I/O related functions:} \= \\
{\tt TSIL\_PrintStatus}	
    \> Print evaluation status to stdout\\
{\tt TSIL\_PrintData} 
    \> Print all integral function values to stdout\\
{\tt TSIL\_WriteData}
    \> Write all integral function values to a file\\
{\tt TSIL\_PrintDataM}
    \> As {\tt TSIL\_PrintData}, but format is valid Mathematica input\\
{\tt TSIL\_WriteDataM}
    \> As {\tt TSIL\_WriteData}, but format is valid Mathematica input\\
{\tt TSIL\_cprintf}
    \> Generic printing of values of {\tt TSIL\_Complex} type\\
{\tt TSIL\_cprintfM}
    \> As {\tt TSIL\_cprintf}, but in Mathematica input form\\
{\tt TSIL\_Error}
    \> Print a message to stderr and exit\\
{\tt TSIL\_Warn}
    \> Print a warning message to stderr\\
{\tt TSIL\_WarnsOff}
    \> Toggles warning messages off. (New in v1.3 June 2015.)\\
{\tt TSIL\_WarnsOn}
    \> Toggles warning messages on. (New in v1.3 June 2015.)\\
{\tt TSIL\_PrintInfo}
    \> Write general information on stdout
\end{tabbing}

\begin{tabbing}
{\tt TSIL\_ResetStepSizeParams} xxx\= \kill
\underline{Utilities:} \> \\
{\tt TSIL\_ResetStepSizeParams}
    \> Sets new parameters used for Runge-Kutta step size control\\
{\tt TSIL\_IsInfinite}
    \> Tests whether argument of type {\tt TSIL\_Complex} is finite \\
{\tt TSIL\_DataSize}
    \> Returns size of intrinsic floating point data used\\
{\tt TSIL\_PrintVersion}
    \> Prints version number of TSIL
\\
{\tt TSIL\_CopyResult}
    \> Copies data from a {\tt TSIL\_DATA} struct to a 
    {\tt TSIL\_RESULT} struct.\\ 
    \> (New in v1.2 July 2014, see end of section 5.2.)\\
{\tt TSIL\_PermuteResult}
    \> Copies the data from one {\tt TSIL\_RESULT} struct to another,
    with \\ \> the option of 
    permuting the squared masses according to either\\ \>
    $(x,z)\leftrightarrow (y,u)$, or
    $(x,y)\leftrightarrow (z,u)$, or $(x,y)\leftrightarrow (u,z)$,
    \\ \> or doing no permutation.
    (New in v1.2 July 2014, see end\\ \> of section 5.2.) \\
{\tt TSIL\_NumFuncs}
    \> Returns number of basis functions of specified type. (New in 
    \\ \> v1.3 June 2015.)
\end{tabbing}

\begin{tabbing}
{\tt TSIL\_Tbaranalytic}xxx\=\kill
\underline{Analytic cases:} \> \\
{\tt TSIL\_Dilog}
    \> Dilogarithm function of complex argument ${\rm Li}_2(z)$\\
{\tt TSIL\_Trilog}    	    
    \> Trilogarithm function of complex argument ${\rm Li}_3(z)$\\
{\tt TSIL\_A}
    \> One-loop vacuum integral $A(x)$\\
{\tt TSIL\_Ap}
    \> One-loop vacuum integral $A(x') = \lnbar(x)$ (New in v1.2 July 2014.)\\
{\tt TSIL\_Aeps}
    \> One-loop vacuum integral $A_\epsilon (x)$\\
{\tt TSIL\_B}
    \> One-loop self-energy integral $B(x,y)$\\
{\tt TSIL\_Bp} 
    \> $B(x',y)$\\
{\tt TSIL\_dBds} 
    \> $\partial B(x,y)/\partial s$\\
{\tt TSIL\_Beps}
    \> One-loop self-energy integral $B_\epsilon(x,y)$\\
{\tt TSIL\_I2}
    \> Two-loop vacuum integral $I(x,y,z)$\\
{\tt TSIL\_I2p}
    \> $I(x',y,z)$\\
{\tt TSIL\_I2p2} 
    \> $I(x'',y,z)$\\
{\tt TSIL\_I2pp} 
    \> $I(x',y',z)$\\
{\tt TSIL\_I2p3}
    \> $I(x''',y,z)$\\
{\tt TSIL\_Sanalytic}
    \> Analytic evaluation of $S$ if available\\
{\tt TSIL\_Tanalytic}
    \> Analytic evaluation of $T$ if available\\
{\tt TSIL\_Tbaranalytic}
    \> Analytic evaluation of $\overline T$ if available\\
{\tt TSIL\_Uanalytic}
    \> Analytic evaluation of $U$ if available\\
{\tt TSIL\_Vanalytic}
    \> Analytic evaluation of $V$ if available\\
{\tt TSIL\_Manalytic}
    \> Analytic evaluation of $M$ if available
\end{tabbing}

\begin{flushleft} 
{\underline{Fortran interface function:}}\\ 
{\tt tsilfevaluate\_}~~~~~~Wrapper for {\tt TSIL\_Evaluate}, 
callable from Fortran
\end{flushleft}

\bigskip \noindent {\it Acknowledgments:} 
We are grateful to M. Kalmykov for bringing to our attention previous 
results on some of the cases in Appendix B.
We also thank 
Nikita Blinov,
Stefano Di Vita,
Philipp Kant,
James McKay,
Andrey Pikelner,
Filippo Sala,
and
Alexander Voigt 
for pointing out bugs in earlier versions.
The work of SPM was supported in 
part by the National Science Foundation grant number PHY-0140129. 
DGR was supported by an award from Research Corporation, and by a grant
from the Ohio Supercomputer Center.

 
\end{document}